\documentclass{article}
\ifdefined\pdfminorversion\pdfminorversion=7\fi

\usepackage[preprint]{neurips_2026}
\usepackage[utf8]{inputenc}
\usepackage[T1]{fontenc}
\usepackage{microtype}
\usepackage{graphicx}
\usepackage{booktabs}
\usepackage{multirow}
\usepackage{adjustbox}
\usepackage[dvipsnames,table]{xcolor}
\usepackage{array}
\usepackage{colortbl}
\usepackage{amsmath,amssymb,amsfonts,mathtools,amsthm}
\usepackage{nicefrac}
\usepackage{caption}
\usepackage{subcaption}
\usepackage{tcolorbox}
\usepackage{soul}
\usepackage{url}
\usepackage{hyperref}
\usepackage[capitalize,noabbrev]{cleveref}
\usepackage{pifont}
\makeatletter
\providecommand{\insert@pcolumn}{\insert@column}
\makeatother
\definecolor{fgrefbg}{RGB}{246,248,252}

\theoremstyle{plain}

\theoremstyle{definition}

\theoremstyle{remark}

\hypersetup{
  hidelinks,
  pdftitle={MulCLIP: A Multi-level Alignment Framework for Enhancing Fine-grained Long-context CLIP},
  pdfauthor={Chau Truong; Hieu Ta Quang; Zhenzhen Liu; Dung D. Le}
}

\title{MulCLIP: A Multi-level Alignment Framework
for Enhancing Fine-grained Long-context CLIP}

\author{
  Chau Truong\textsuperscript{1,*} \quad
  Hieu Ta Quang\textsuperscript{2,*} \quad
  Zhenzhen Liu\textsuperscript{3,\textdagger} \quad
  Dung D. Le\textsuperscript{2,\textdagger}
  \\[0.5em]
  \textsuperscript{1}Vinmotion, Hanoi, Vietnam \\
  \textsuperscript{2}VinUniversity, Hanoi, Vietnam \\
  \textsuperscript{3}Cornell University
}

\begin{document}

\maketitle

\begingroup
\renewcommand{\thefootnote}{\fnsymbol{footnote}}
\footnotetext[1]{Equal technical contribution.}
\footnotetext[2]{Equal supervision. Correspondence: \texttt{dung.ld@vinuni.edu}.}
\endgroup

\begin{abstract}
  Vision-language models such as CLIP excel at image-text alignment but struggle with long, detailed descriptions due to training on short captions. Recent methods address this limitation using region proposals to align visual regions with sentences, but at substantial deployment cost. We present MulCLIP, an end-to-end multi-level alignment framework that directly exploits natural long-text structure without region proposals. Instead of simply stacking objectives, MulCLIP aligns different textual granularities through compatible training signals: global contrastive alignment for long captions and short summaries, Word--Patch Reconstruction over locally calibrated features for within-sample word--patch semantics, and Subcaption--Aggregated Patch alignment for context-rich subcaption grounding. Experiments on benchmarks spanning varying caption lengths show consistent gains, and ablations confirm that the proposed objectives are complementary and jointly improve the overall framework. Code will be released at \url{https://github.com/truongchau2602/MulCLIP}.
\end{abstract}

\section{Introduction}
Bridging the alignment gap between vision and language is a central goal of vision-language representation learning. CLIP~\citep{refCLIP} tackles this with a contrastive dual-encoder trained on large-scale image-text pairs, enabling strong zero-shot transfer. However, CLIP struggles with fine-grained understanding of attributes, relations, and compositional descriptions~\citep{refLotLIP, refEyewideshut}, largely because it is trained mostly on short, generic captions and relies on a global image-text objective that does not explicitly supervise localized grounding. These limitations become more severe in complex scenes and long-form descriptions that require correspondences across multiple textual granularities.

To address this issue, LongCLIP \citep{refLongCLIP} extends CLIP's capacity for long-text modeling by modifying its positional encodings, enabling the model to process longer sequences without disrupting the alignment learned from pre-trained CLIP weights.  While effective, it still operates at the global representation level and fails to capture the fine-grained correspondences that naturally arise in detailed descriptions.
FineLIP~\citep{refFineLIP} narrows this gap by refining visual and textual tokens and aligning them through a Cross-Modal Late Interaction Module (CLIM), which aggregates bidirectional token-wise maximum similarities into an image--text matching score. However, it focuses solely on long captions, overlooking the semantically rich short phrases describing specific image regions \citep{refDOCCI, refDCI}. In addition, training only on long captions can degrade short-caption understanding, as demonstrated in prior findings of \citet{refLotLIP}. GOAL \citep{refGOAL} tackles both long and sub captions via a weakly supervised global--local alignment, mining pseudo region--sentence pairs from SAM-based crops and similarity CLIP-scored matching. While it achieves significant fine-grained performance, its reliance on external segmentation and post-hoc filtering increases compute and limits deployment flexibility.
\begin{figure}[t]
\begin{center}
    \includegraphics[width=\columnwidth]{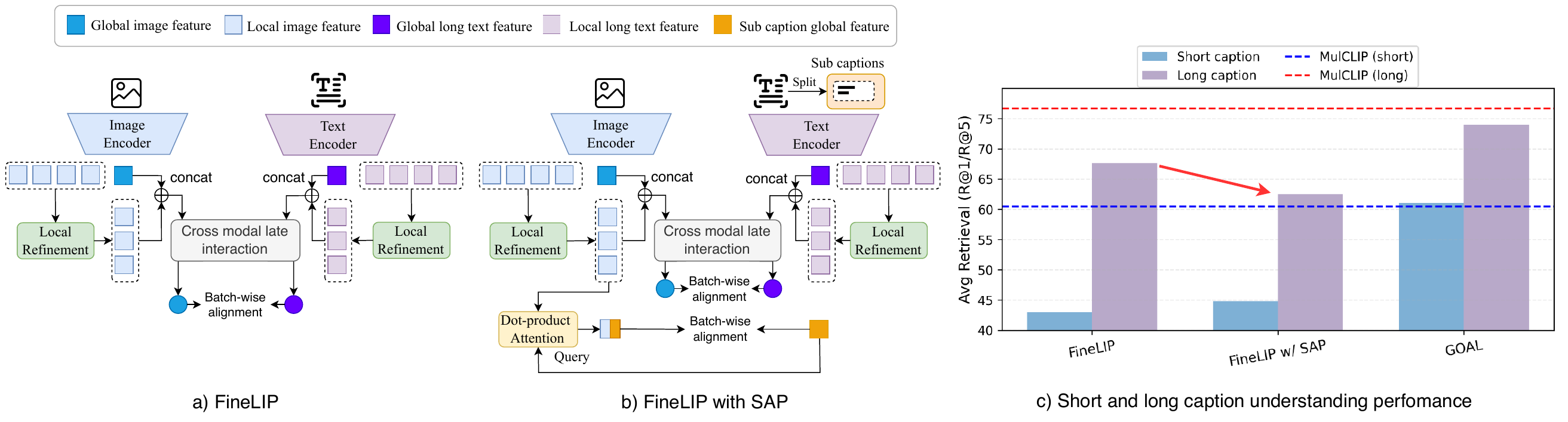}
\end{center}
\caption{\textbf{The major challenge in unified long-context learning.} Subcaption-patched self supervision cannot be easily incorporated into existing FineLIP training - This provides only marginal improvements in short-text understanding performance, while drastically degrading long-text understanding.}
\label{fig:challenge}
\end{figure}

In this work, we aim to design a unified adaptation framework that bridges long-form descriptions and complex visual content across multiple textual granularities without relying on any external modules. However, simply stacking objectives across granularities is not sufficient. In FineLIP-style designs, global and refined local tokens are combined and jointly aligned through CLIM~\citep{refFineLIP, refFILIP}. This coupling makes it difficult to incorporate subcaption-level self-supervision as a clean additional signal, since the new sentence-level regional objective is optimized through the same token space used for long-caption late interaction. As shown in Fig. \ref{fig:challenge}, directly adding a subcaption--patch objective on top of such a global-local late-interaction baseline can noticeably reduce long-caption understanding. Further ablation analysis indicates that this gap is associated with the choice of token-level objective. This suggests that multi-level objectives do not always compose well when optimized on the same coupled token space, and that subcaption-level grounding requires a more compatible word-level training signal.

\begin{figure}[t]
\begin{center}
    \includegraphics[width=0.83\columnwidth]
    {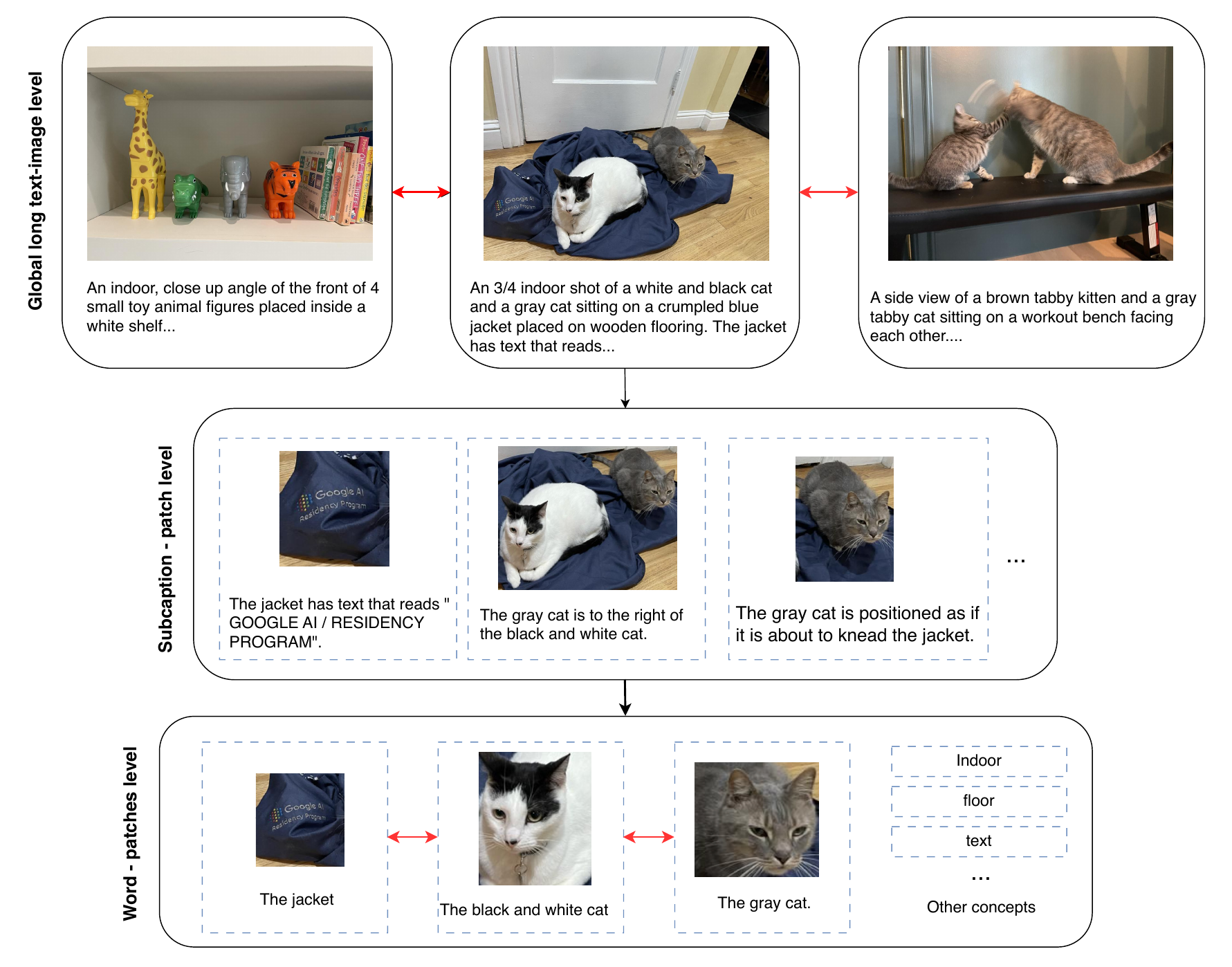}
\end{center}
\caption{\textbf{Multi-granularity alignment logic for long-context understanding.}
We apply batch-wise contrastive alignment at the global caption--image and subcaption--patch levels, and within-sample contrastive alignment at the word/phrase--patch level to learn localized correspondences.}
\label{fig:multi-granularity}
\end{figure}

This motivates a simple design principle. As illustrated in Fig. \ref{fig:multi-granularity}, long captions mix different semantic units. Words or phrases usually point to a single entity or attribute, while subcaptions often describe broader facts---sometimes involving multiple objects and shared visual evidence. Because of this overlap, subcaption-level training is better served by batch-wise contrastive, whereas word-level grounding is more localized and fits naturally with within-sample contrastive.
Our main contributions are summarized as follows:

\textbullet{} We identify a compatibility issue in existing late-interaction designs, where the effectiveness of adding subcaption-level supervision depends on the token-level alignment objective.

\textbullet{} We propose MulCLIP, a proposal-free multi-granularity framework that decouples caption-, subcaption-, and word-level alignment through compatible objectives.

\textbullet{} We validate MulCLIP on long- and short-caption retrieval benchmarks, showing improved long-context understanding while preserving short-text transfer, with ablations confirming the complementarity of its components.
\section{Related Work}

\textbf{Vision-Language Models (VLMs).} Contrastive learning has established itself as a leading paradigm for multimodal pre-training, significantly advancing the field of image-text alignment. The pioneering work of CLIP \citep{refCLIP}, employing a dual-encoder architecture trained contrastively on approximately 400 million image--caption pairs, demonstrates robust zero-shot transfer capabilities across various downstream tasks such as open-vocabulary recognition, object detection, and semantic segmentation. Moreover, CLIP has become an essential component in numerous generative vision--language systems, including multimodal language models like LLaVA \citep{refLLAVA} and diffusion models \citep{refGLIDE, refStableDiffusion}. Following CLIP's success, the next VLM foundation models train on hundreds of millions to billions of image-text pairs \citep{refALIGN, refBLIP} and this trend also propagates into domain-specific VLMs, such as medical imaging application \citep{refBiomedCLIP}. However, these models typically rely on short, broad image descriptions as captions, causing them to miss crucial local-level detailed information.

\textbf{Fine-grained understanding in VLMs.} To address these limitations, recent work targets fine-grained attributes in long text. Some approaches integrate the inherent short descriptions from synthetic long text to vision-language models and retrained it from scratch \citep{refDreamLIP, refLotLIP, refFLAIR}, but this forfeits the rich knowledge of pre-trained models like CLIP, demands large-scale data and computation. Moreover, \citet{refRevisit} show that pre-training CLIP solely on synthetic captions can substantially degrade performance, despite their stronger image-text alignment.

An alternative, more efficient approach involves fine-tuning existing pre-trained CLIP model. Early works \citep{refGLoRIA, refSPARC}, highlight token-level alignment between image patches and text word embeddings, pushing the boundaries of fine-grained image--text understanding. In an emerging direction, LongCLIP \citep{refLongCLIP} or TULIP \citep{refTulip} extends the token capacity of CLIP's text encoder, enabling it to process and represent longer, more descriptive captions. FG-CLIP \citep{refFG_CLIP} augments CLIP-400M with 1.6B LLM-generated image-text samples. However, heavy reliance on synthetic captions may propagate captioner biases and hallucination noise.

In addition, several dense, detailed image-caption datasets such as DCI \citep{refDCI} and DOCCI \citep{refDOCCI} have been introduced, leveraging large vision--language models (LVLMs) and human annotation to generate high quality sub-captions that describe local visual details. Recent approach such as GOAL \citep{refGOAL} exploits such annotations by using SAM to mine pseudo region--subcaption pairs, followed by CLIP-based filtering and joint global--local alignment. FineLIP \citep{refFineLIP} instead refines visual and textual tokens and applies cross-modal late interaction for long-caption alignment. However, all of these approaches are either non-unified or address at most two granularities, leaving the gap of unified and effective alignment strategy for fine-grained long-context learning.

\section{Method}
\label{sec:method}
\subsection{Global-level Alignment}
To handle text sequences longer than CLIP's standard 77 token limit, we first adopt LongCLIP's positional embedding interpolation strategy in our text encoder. This adjustment allows longer text inputs while minimizing disruptions to the strong crossmodal alignment achieved in the pretrained CLIP.

Consider a CLIP-style vision--language model \( f = (f_v, f^h_v, f_t, f^h_t) \), where \(f_v\) and \(f_t\) denote image and text backbone modules respectively, and \(f^h_v\) and \(f^h_t\) represent corresponding projection heads mapping embeddings to a shared \(d\)-dimensional space. Given an image \(I\) and its associated long-form caption \(T_{long}\), we first segment \(T_{long}\) into \(M\) sentence-level subcaptions \(\{T_{sub}^{i}\}_{i=1}^{M}\). We then extract the image's global and local features using \(f_v\) and project them using \(f^h_v\):
\begin{equation}
    \left[v_{cls},\;v_{loc} \right] = f^h_v(f_v(I)) \in \mathbb{R}^{(P+1) \times d},
\end{equation}
where \(v_{cls}\in \mathbb{R} ^{d}\) denotes the global [CLS] embedding of an image and \(v_{loc}\in \mathbb{R} ^{P\times d}\) are \(P\) patch local embeddings.

Similarly, text embeddings are obtained from the text encoder:
\begin{equation}
\begin{alignedat}{2}
[t^{long}_{eot}, t^{long}_{loc}]
&= f^h_t(f_t(T_{long})) \in \mathbb{R}^{(K+1)\times d},
\quad
[\{t^{sub}_{eot,i}\}_{i=1}^{M}, \cdot]
&= f^h_t(f_t(\{T^i_{sub}\}_{i=1}^{M})) \in \mathbb{R}^{M\times d}.
\end{alignedat}
\label{eq:sub_embedding}
\end{equation}

where \(t^{long}_{eot}\in \mathbb{R} ^{d}\) and \(t^{long}_{loc}\in \mathbb{R}^{K \times d}\) denotes the global [EOT] and K local embeddings of long text, while \( t^{sub}_{eot}= \{t_{eot,\text{i}}^{\text{sub}}\}_{i=1}^{M} \in \mathbb{R}^{M \times d}\) denotes the global embeddings of \(M\) subcaptions.

During training, each image is globally aligned with both its long caption and short summary caption, enabling long-context adaptation while preserving short-text understanding. We define the global objective as:
\begin{equation}
    \mathcal{L}_{\text{global}}
    = \mathcal{L}^{\text{batch}}_{\text{contrast}}(v_{cls},\; t^{long}_{eot})
    + \tfrac{1}{2}\,\mathcal{L}^{\text{batch}}_{\text{contrast}}(v_{cls},\; t^{short}_{eot})
\end{equation}
where \(\mathcal{L}^{\text{batch}}_{\text{contrast}}\) is a batch-level contrastive loss.
\begin{figure*}[t]
    \centering  \includegraphics[width=0.8\textwidth]
    {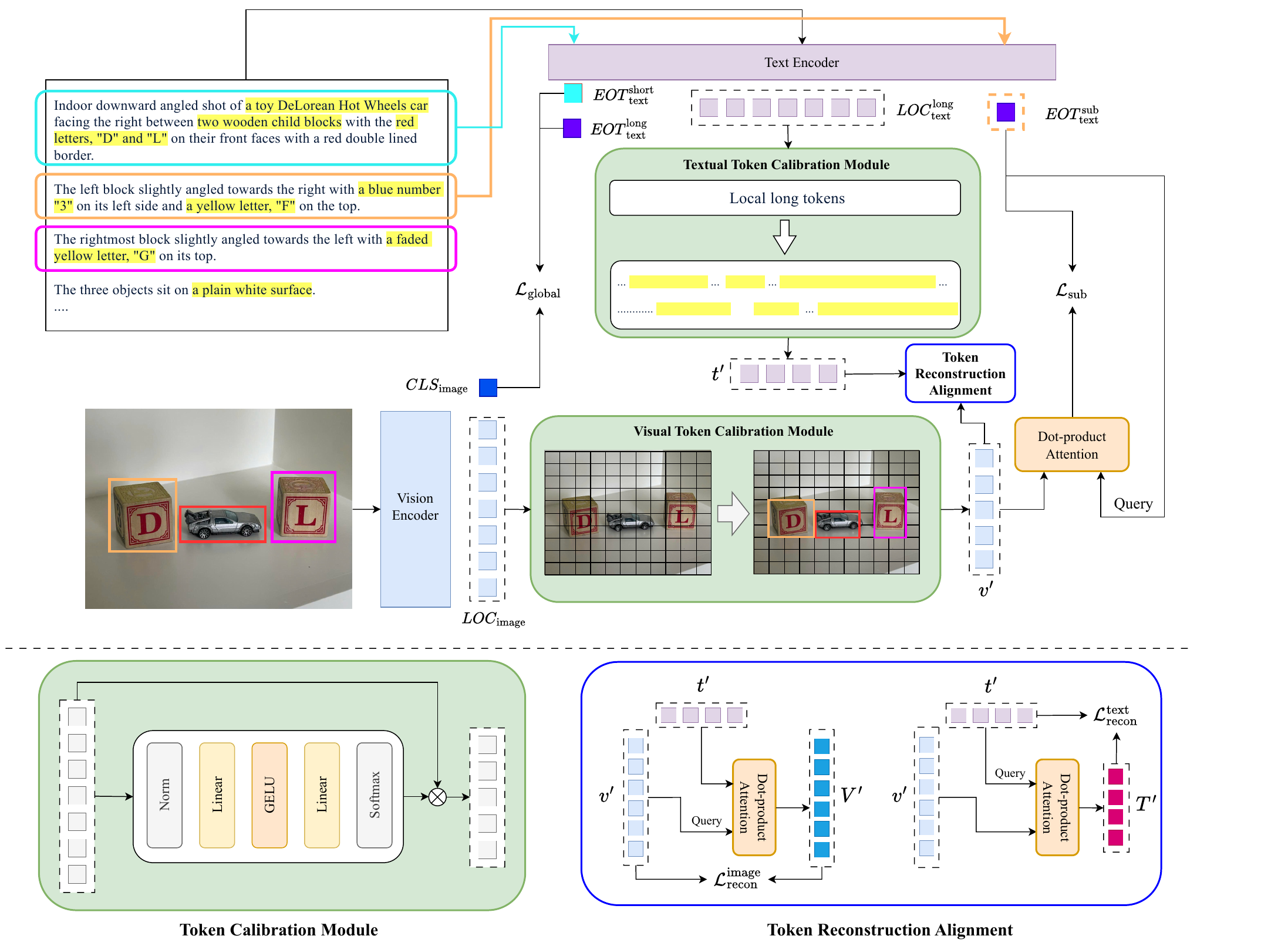}
     \caption{\textbf{Overview of MulCLIP}.
    Following the multi-granularity principle in Fig.~\ref{fig:multi-granularity}, MulCLIP aligns images and long-form text through compatible objectives. It preserves global alignment with long captions and short summaries, refines local visual and textual tokens, learns word--patch correspondences via WPR with within-sample contrastive learning, and grounds subcaptions via SAP with batch-wise contrastive learning.}
    \vspace{-6pt}
    \label{fig:pipeline}
    \vspace{-0.2cm}
\end{figure*}

\subsection{Fine-grained Cross-modal Alignment}
\paragraph{Local Token Calibration.}
In dense and highly-aligned image-text pairs, redundancy and ambiguity frequently occur in both local image patches \citep{refLAPS, refTOME} and local long-text tokens \citep{refFineLIP}. To mitigate these issues, we adopt aggregation network \citep{refSlim_VIT}, as adaptive calibration mechanism for both visual and textual local embeddings. Specifically, given an input sequence of \(N\) tokens, each with dimension \(d\), we denote the input as \(X \in \mathbb{R}^{N \times d}\). The aggregated output \(X' \in \mathbb{R}^{N' \times d}\) is computed as:
\begin{align}
X'=\mathrm{SoftMax}\!\Bigl(\frac{W_q\,\sigma(XW_k)^\top}{\tau}\Bigr)X,
\end{align}
where \(W_k \in \mathbb{R}^{d \times d_k}\) and \(W_q \in \mathbb{R}^{N' \times d_k}\) are learnable projection matrices (\( d_k< d\)), which \(N'/N=0.5\) by default, \(\sigma(\cdot)\) is a non-linear activation GELU \citep{gelu}, and \(\tau\) is a learnable temperature parameter. We apply calibration modules independently to visual patches \(v_{loc}\in \mathbb{R} ^{P\times d}\) and local long caption tokens \(t^{long}_{loc}\in \mathbb{R}^{K \times d}\), yielding refined patches \(v'=\{\tilde{v}_i\}^{rP}_{i=1}\!\in\!\mathbb{R}^{rP\times d}\) and refined words \(t'=\{\tilde{t}_i\}^{rK}_{i=1} \!\in\!\mathbb{R}^{rK\times d}\). To further leverage these semantic tokens for fine-grained matching, we propose two complementary alignment strategies: \textit{token reconstruction alignment} and \textit{subcaption--aggregated patch alignment} that operate on top of them.
\paragraph{Token Reconstruction Alignment.} To align semantic words with their corresponding visual patches, we use the reduced sequences \(v^{\prime}\) and \(t^{\prime}\) as queries in a bidirectional dot-product attention:
\begin{equation}
\begin{alignedat}{2}
A_{v \to t} &= \mathrm{SoftMax}\!\left(\frac{v^{\prime}(t^{\prime})^\top}{\sqrt{d}}\right),
\qquad
A_{t \to v} &= \mathrm{SoftMax}\!\left(\frac{t^{\prime}(v^{\prime})^\top}{\sqrt{d}}\right).
\end{alignedat}
\end{equation}
These matrices select, for every image patch, the most relevant text token and vice-versa, yielding cross-modal reconstructions \(V^{\prime} \in \mathbb{R} ^{(rP) \times d}\) and \(T^{\prime} \in \mathbb{R} ^{(rK) \times d} \) :
\begin{equation}
V^{\prime}= \{\tilde{V}_i\}_{i=1}^{rP} = A_{v \to t}t^{\prime},
\quad
T^{\prime}= \{\tilde{T}_i\}_{i=1}^{rK} = A_{t \to v}v^{\prime}.
\end{equation}
We introduce a self-sample alignment objective that applies two contrastive terms, one for images and one for text, to make every refined token consistent with its cross-modal reconstruction. Specifically, we impose contrastive losses for each token within the same sample; therefore, no cross-sample negatives are needed. The reconstruction losses are computed as:
\begin{equation}
\begin{alignedat}{2}
\mathcal{L}_{\text{recon}}^{\text{image}}(v', V')
&= \frac{1}{rP}\sum_{i=1}^{rP}
\mathcal{L}^{\text{sample}}_{\text{contrast}}(\tilde{v}_i,\tilde{V}_i),
\quad
\mathcal{L}_{\text{recon}}^{\text{text}}(t', T')
&= \frac{1}{rK}\sum_{i=1}^{rK}
\mathcal{L}^{\text{sample}}_{\text{contrast}}(\tilde{t}_i,\tilde{T}_i).
\end{alignedat}
\end{equation}
The final objective of \textbf{W}ord-\textbf{P}atch \textbf{R}econstruction (\textbf{WPR}) process is simply the sum:
\begin{align}
\mathcal{L}_{\text{Word}}(v', t')
= \mathcal{L}_{\text{recon}}^{\text{image}}(v', V')
+ \mathcal{L}_{\text{recon}}^{\text{text}}(t', T')
\end{align}
\paragraph{Subcaption--Aggregated Patch Alignment.}
Descriptive captions from curated long-text image datasets typically consist of multiple sentence-level subcaptions, each of which may describe a local image region. To explicitly align these subcaptions with corresponding visual content, we obtain each subcaption embedding \(t_{eot, \text{i}}^{\text{sub}}\in \mathbb{R}^{1 \times d}\) from Eq.\ref{eq:sub_embedding} and associate them with the aggregated visual representation from the refined local embeddings \(v'\in\!\mathbb{R}^{rP\times d}\). Specifically, we use attention weights derived from dot-product similarity between each subcaption embedding and the visual patches:
\begin{align}
\bar{v}^i = \mathrm{SoftMax}\left(\frac{t_{eot, \ \text{i}}^{\text{sub}}(v^{\prime})^\top}{\sqrt{d}}\right)  v^{\prime} \in \mathbb{R}^{d}
\end{align}
We then impose a \textbf{S}ubcaption-\textbf{A}ggregated \textbf{P}atch (\textbf{SAP}) objective that applies a contrastive loss between each subcaption embedding and its aggregated visual representation:
\begin{align}
\mathcal{L}_{\text{Sub}}(v', t^{sub}_{eot}) = \frac{1}{M}\sum_{i=1}^{M}\mathcal{L}^{\text{batch}}_{\text{contrast}}(\bar{v}^{i}, t^{sub}_{eot,i}),
\end{align}
\subsection{Overall Alignment Objective.}
To enable robust and comprehensive vision--language alignment, we jointly optimize three complementary objectives:
\begin{equation}
\mathcal{L}_{total} = \mathcal{L}_{\mathrm{global}} + \lambda_W \mathcal{L}_{\mathrm{Word}}(v',t') +  \lambda_S \mathcal{L}_{\mathrm{Sub}}(v', t^{sub}_{eot}),
\end{equation}
where $\lambda_W$ and $\lambda_S$ are hyperparameters set to 1, so that each loss term operates on a similar scale. We employ a sigmoid-based contrastive loss \citep{refSiglip} as the main objective for all $\mathcal{L}_{\text{contrast}}$ terms.

\section{Experiments}
\label{sec:exp}
\subsection{Experimental Setup}
\paragraph{Datasets.} We fine-tune MulCLIP on the training splits of both DOCCI and DCI \citep{refDOCCI, refDCI} and evaluate on the 3 test sets of DOCCI, DCI and Urban1K \citep{refLongCLIP} to measure both in-domain and out-of-domain fine-grained long-text retrieval performance comprehensively on Table \ref{tab:cross_dci_docci}. The DOCCI dataset comprises 9,647 training examples and a test split totaling 5,100 samples (5,000 from the official test set plus 100 from the qualification set). Since DCI's original test split contains only 100 examples,
we follow \citep{refGOAL} and randomly hold out 2,000 examples from its 7,805-example training pool for evaluation, using the remaining 5,805 examples for fine-tuning. The held-out examples are excluded from training, and all compared methods use the same split. To compare the short-text performance, we evaluate our model on the validation set of COCO2017 \citep{refCOCO} and Flickr30k \citep{refFlick30k}.

\paragraph{Training setting.}
To validate our approach, we fine-tune two CLIP variants, ViT-B/16 and ViT-L/14, for 8 epochs, using a batch size of 16 for both ViT-B/16 and ViT-L/14. We set the base backbone learning rate to \(1\times10^{-5}\) and the refinement-module learning rate to \(2\times10^{-4}\), so as to retain the pre-trained CLIP representations while encouraging the refinement layers to adapt to our long-caption datasets. More details of training setting are provided in the supplementary material Sec.~\ref{app:implementation}.
\paragraph{Test settings and state-of-the-art comparisons}
 We measure Text-to-Image (T2I) and Image-to-Text (I2T) retrieval performance using Recall@\emph{k}. We compare MulCLIP against leading adaptation methods tailored for downstream long-caption datasets, such as FineLIP and GOAL. For reference, we additionally report \textbf{FG-CLIP}, a large-scale continued CLIP pretraining baseline trained on 1.64B image-text pairs, which represents a different regime.

\subsection{Results}

\begin{table*}[t]
  \centering
  \caption{\textbf{Long-text retrieval performance on DOCCI, DCI and Urban1k}. Rows (DOCCI FT, DCI FT) indicate the dataset on which the methods were trained. We highlight the best performance (blue shading) and \underline{second-best} among the adaptation methods within each backbone; FG-CLIP is a reference and excluded from ranking. Gray shading indicates in-domain retrieval (diagonal blocks). \textbf{Avg@1} is the mean of all R@1 values in the same row.}
  \label{tab:cross_dci_docci}
  \adjustbox{width=\textwidth,center}{
    \begin{tabular}{@{} l l l
        c c c c
        c c c c
        c c c c
        c @{}}
      \toprule
       &  & \textbf{Method}
       & \multicolumn{4}{c}{\textbf{DOCCI}}
       & \multicolumn{4}{c}{\textbf{DCI}}
       & \multicolumn{4}{c}{\textbf{Urban1k}} \\
      \cmidrule(lr){4-7}\cmidrule(lr){8-11}\cmidrule(lr){12-15}
       &  &
       & \multicolumn{2}{c}{\textbf{Text-to-Image}}
       & \multicolumn{2}{c}{\textbf{Image-to-Text}}
       & \multicolumn{2}{c}{\textbf{Text-to-Image}}
       & \multicolumn{2}{c}{\textbf{Image-to-Text}}
       & \multicolumn{2}{c}{\textbf{Text-to-Image}}
       & \multicolumn{2}{c}{\textbf{Image-to-Text}}
       & \textbf{Avg@1}\\
      \cmidrule(lr){4-5}\cmidrule(lr){6-7}
      \cmidrule(lr){8-9}\cmidrule(lr){10-11}
      \cmidrule(lr){12-13}\cmidrule(lr){14-15}
       &  &
       & R@1 & R@5 & R@1 & R@5
       & R@1 & R@5 & R@1 & R@5
       & R@1 & R@5 & R@1 & R@5 & \\
      \midrule
      
        & \scriptsize B/16
        & FG-CLIP
          & \textcolor{gray!60}{84.80} & \textcolor{gray!60}{97.75}
          & \textcolor{gray!60}{84.06} & \textcolor{gray!60}{97.29}
          & \textcolor{gray!60}{75.64} & \textcolor{gray!60}{89.64}
          & \textcolor{gray!60}{75.74} & \textcolor{gray!60}{89.34}
          & \textcolor{gray!60}{93.90} & \textcolor{gray!60}{99.10}
          & \textcolor{gray!60}{94.00} & \textcolor{gray!60}{99.60}
          & \textcolor{gray!60}{84.69} \\
        \cmidrule(lr){2-16}
        & \scriptsize L/14
        & FG-CLIP
          & \textcolor{gray!60}{89.67} & \textcolor{gray!60}{98.73}
          & \textcolor{gray!60}{87.75} & \textcolor{gray!60}{98.37}
          & \textcolor{gray!60}{80.79} & \textcolor{gray!60}{91.15}
          & \textcolor{gray!60}{80.49} & \textcolor{gray!60}{91.80}
          & \textcolor{gray!60}{96.60} & \textcolor{gray!60}{99.80}
          & \textcolor{gray!60}{96.60} & \textcolor{gray!60}{99.90}
          & \textcolor{gray!60}{88.65} \\
      \midrule
      \multirow{6}{*}{\rotatebox{90}{\scriptsize DOCCI FT}}
        & \multirow{3}{*}{\rotatebox{90}{\scriptsize ViT-B/16}}
           & FineLip
              & \cellcolor{gray!15}70.94 & \cellcolor{gray!15}92.98
              & \cellcolor{gray!15}71.50 & \cellcolor{gray!15}93.24
              & 58.58 & 78.64 & 59.88 & 80.39
              & 67.50 & 88.00 & 77.40 & 93.90
              & 67.63 \\
        &   & GOAL
              & \cellcolor{gray!15}\underline{79.47} & \cellcolor{gray!15}\underline{96.65}
              & \cellcolor{gray!15}\underline{79.43} & \cellcolor{gray!15}\underline{96.14}
              & \underline{64.13} & \underline{82.69} & \underline{65.88} & \underline{83.44}
              & \underline{73.20} & \cellcolor{blue!15}92.70
              & \underline{81.90} & \underline{95.80}
              & \underline{74.00} \\
        &   & \textbf{MulCLIP}
              & \cellcolor{blue!15}82.2  & \cellcolor{blue!15}97.12
              & \cellcolor{blue!15}80.26 & \cellcolor{blue!15}96.88
              & \cellcolor{blue!15}69.08 & \cellcolor{blue!15}85.99
              & \cellcolor{blue!15}67.13 & \cellcolor{blue!15}84.24
              & \cellcolor{blue!15}77.30 & \underline{92.60}
              & \cellcolor{blue!15}84.00 & \cellcolor{blue!15}96.10
              & \cellcolor{blue!15}76.66 \\
        \cmidrule(lr){2-16}
        & \multirow{3}{*}{\rotatebox{90}{\scriptsize ViT-L/14}}
           & FineLip
              & \cellcolor{gray!15}74.70 & \cellcolor{gray!15}94.24
              & \cellcolor{gray!15}75.44 & \cellcolor{gray!15}94.60
              & 62.88 & 81.69 & 63.68 & 83.44
              & 67.40 & 87.60 & 78.70 & 94.20
              & 70.47 \\
        &   & GOAL
              & \cellcolor{gray!15}\underline{84.37} & \cellcolor{blue!15}99.55
              & \cellcolor{gray!15}\underline{82.57} & \cellcolor{gray!15}\underline{97.37}
              & \underline{68.93} & \underline{85.74}
              & \underline{68.43} & \underline{85.99}
              & \underline{83.00} & \underline{95.40}
              & \underline{86.30} & \underline{96.50}
              & \underline{78.93} \\
        &   & \textbf{MulCLIP}
              & \cellcolor{blue!15}86.73 & \cellcolor{gray!15}\underline{98.10}
              & \cellcolor{blue!15}84.80 & \cellcolor{blue!15}97.88
              & \cellcolor{blue!15}72.93 & \cellcolor{blue!15}88.00
              & \cellcolor{blue!15}72.03 & \cellcolor{blue!15}86.64
              & \cellcolor{blue!15}85.80 & \cellcolor{blue!15}97.10
              & \cellcolor{blue!15}88.30 & \cellcolor{blue!15}97.30
              & \cellcolor{blue!15}81.77 \\
      \midrule
      \multirow{6}{*}{\rotatebox{90}{\scriptsize DCI FT}}
        & \multirow{3}{*}{\rotatebox{90}{\scriptsize ViT-B/16}}
           & Finelip
              & 65.50 & 89.30 & 66.32 & 90.72
              & \cellcolor{gray!15}60.38 & \cellcolor{gray!15}80.39
              & \cellcolor{gray!15}63.58 & \cellcolor{gray!15}82.94
              & 64.00 & 84.60 & 78.60 & 94.90
              & 66.40 \\
        &   & GOAL
              & \underline{71.22} & \underline{92.39}
              & \underline{71.18} & \underline{92.88}
              & \cellcolor{gray!15}\underline{72.64} & \cellcolor{blue!15}89.89
              & \cellcolor{blue!15}72.84 & \cellcolor{blue!15}90.50
              & \underline{77.20} & \underline{93.70}
              & \underline{82.90} & \underline{96.80}
              & \underline{74.66} \\
        &   & \textbf{MulCLIP}
              & \cellcolor{blue!15}73.78 & \cellcolor{blue!15}93.86
              & \cellcolor{blue!15}71.75 & \cellcolor{blue!15}92.96
              & \cellcolor{blue!15}75.13 & \cellcolor{gray!15}\underline{89.44}
              & \cellcolor{gray!15}\underline{72.00} & \cellcolor{gray!15}\underline{89.24}
              & \cellcolor{blue!15}80.90 & \cellcolor{blue!15}93.90
              & \cellcolor{blue!15}85.20 & \cellcolor{blue!15}97.00
              & \cellcolor{blue!15}76.46 \\
        \cmidrule(lr){2-16}
        & \multirow{3}{*}{\rotatebox{90}{\scriptsize ViT-L/14}}
           & FineLip
              & 68.84 & 90.92 & 71.54 & 92.56
              & \cellcolor{gray!15}66.03 & \cellcolor{gray!15}84.49
              & \cellcolor{gray!15}65.58 & \cellcolor{gray!15}85.19
              & 68.50 & 86.10 & 79.50 & 94.80
              & 70.00 \\
        &   & GOAL
              & \underline{79.04} & \underline{95.78}
              & \cellcolor{blue!15}79.16 & \cellcolor{blue!15}95.96
              & \cellcolor{gray!15}\underline{76.89} & \cellcolor{gray!15}\underline{91.05}
              & \cellcolor{gray!15}\underline{76.59} & \cellcolor{gray!15}\underline{91.20}
              & \underline{84.50} & \underline{96.40}
              & \cellcolor{blue!15}89.80 & \underline{97.80}
              & \underline{81.00} \\
        &   & \textbf{MulCLIP}
              & \cellcolor{blue!15}81.04 & \cellcolor{blue!15}96.33
              & \underline{78.35} & \underline{95.31}
              & \cellcolor{blue!15}78.83 & \cellcolor{blue!15}91.39
              & \cellcolor{blue!15}76.83 & \cellcolor{blue!15}92.09
              & \cellcolor{blue!15}88.10 & \cellcolor{blue!15}97.00
              & \underline{89.70} & \cellcolor{blue!15}97.90
              & \cellcolor{blue!15}82.14 \\

      \bottomrule
    \end{tabular}
  }
\end{table*}

\begin{table*}[ht!]
  \centering
  \caption{\textbf{Zero-shot short caption retrieval on COCO and Flickr30k}. MulCLIP shows competitive performance, often matching or exceeding GOAL across different metrics and model backbones.}
  \label{tab:short_text_retrieval}
  \resizebox{0.75\linewidth}{!}{%
    \begin{tabular}{@{} l l l
                     c c c c
                     c c c c
                     c @{}}
      \toprule
       &  & \multirow{2}{*}{\textbf{Method}}
       & \multicolumn{4}{c}{\textbf{COCO}}
       & \multicolumn{4}{c}{\textbf{Flickr30k}} \\
      \cmidrule(lr){4-7}\cmidrule(lr){8-11}
       &  &
       & \multicolumn{2}{c}{\textbf{Text-to-Image}} & \multicolumn{2}{c}{\textbf{Image-to-Text}}
       & \multicolumn{2}{c}{\textbf{Text-to-Image}} & \multicolumn{2}{c}{\textbf{Image-to-Text}}
       & \textbf{Avg@1}  \\
      \cmidrule(lr){4-5}\cmidrule(lr){6-7}\cmidrule(lr){8-9}\cmidrule(lr){10-11}
       &  &
       & R@1 & R@5 & R@1 & R@5 & R@1 & R@5 & R@1 & R@5 &  \\
      \midrule
       & \multirow{8}{*}{\rotatebox{90}{\scriptsize ViT-B/16}}
       & FG-CLIP
           & \textcolor{gray!60}{43.75} & \textcolor{gray!60}{69.46} & \textcolor{gray!60}{62.48} & \textcolor{gray!60}{84.20}
           & \textcolor{gray!60}{74.38} & \textcolor{gray!60}{92.10} & \textcolor{gray!60}{89.50} & \textcolor{gray!60}{99.00} & \textcolor{gray!60}{67.53} \\
       \addlinespace[2pt]\cmidrule(lr){3-12}
         & & Zeroshot CLIP
           & 33.95 & 59.46 & 54.14 & 77.74
           & 63.20 & 86.30 & \underline{82.90} & \cellcolor{blue!15}97.20 & 58.55 \\

       &  & FineLIP DOCCI FT
           & 36.30 & 61.77 & \underline{56.68} & \underline{80.14}
           & 29.93 & 53.63 & 49.11 & 72.71 & 43.00 \\
       &  & GOAL DOCCI FT
           & \underline{37.28} & \underline{62.96} & \cellcolor{blue!15}56.84 & \cellcolor{blue!15}80.20
           & \underline{66.92} & \underline{88.56} & \cellcolor{blue!15}83.20 & \underline{96.70} & \cellcolor{blue!15}61.06 \\
       &  & \textbf{MulCLIP DOCCI FT}
           & \cellcolor{blue!15}{37.68} & \cellcolor{blue!15}63.26 & 54.76 & 78.64
           & \cellcolor{blue!15}67.44 & \cellcolor{blue!15}88.98 & 81.90 & 96.30 & \underline{60.45} \\
       \addlinespace[3pt]\cmidrule(lr){3-12}
       &  & FineLIP DCI FT
           & 35.44 & 61.18 & \underline{55.48} & \cellcolor{blue!15}79.38
           & 29.07 & 53.24 & 48.43 & 72.64 & 42.11 \\
       &  & GOAL DCI FT
           & \underline{37.20} & \underline{63.17} & \cellcolor{blue!15} 55.82 & \underline{79.10}
           & \underline{66.12} & \underline{88.42} & \underline{82.70} & \cellcolor{blue!15}{96.60} & \underline{60.46} \\
       &  & \textbf{MulCLIP DCI FT}
           & \cellcolor{blue!15}37.69 & \cellcolor{blue!15}63.34 & 53.84 & 78.00
           & \cellcolor{blue!15}{67.34} & \cellcolor{blue!15}88.98 & \cellcolor{blue!15}{83.00} & \underline{96.50} & \cellcolor{blue!15}60.47 \\
      \midrule
       & \multirow{8}{*}{\rotatebox{90}{\scriptsize ViT-L/14}}
       & FG-CLIP
           & \textcolor{gray!60}{50.26} & \textcolor{gray!60}{74.48} & \textcolor{gray!60}{68.78} & \textcolor{gray!60}{87.78}
           & \textcolor{gray!60}{80.50} & \textcolor{gray!60}{95.00} & \textcolor{gray!60}{93.80} & \textcolor{gray!60}{99.60} & \textcolor{gray!60}{73.34} \\
       \addlinespace[2pt]\cmidrule(lr){3-12}
         & & Zeroshot CLIP
           & 37.29 & 61.82 & 57.68 & 80.20
           & 65.38 & 87.36 & 86.70 & 94.50 & 61.76 \\

       &  & FineLIP DOCCI FT
           & 41.18 & 65.96 & 59.14 & 82.00
           & 36.66 & 60.33 & 53.49 & 77.59 & 47.62 \\
       &  & GOAL DOCCI FT
           & \cellcolor{blue!15}44.22 & \underline{69.19} & \cellcolor{blue!15}62.82 & \cellcolor{blue!15}84.04
           & \underline{73.88} & \underline{92.22} & \cellcolor{blue!15}89.80 & \cellcolor{blue!15}98.60 & \cellcolor{blue!15}67.68 \\
       &  & \textbf{MulCLIP DOCCI FT}
           & \underline{43.69} & \cellcolor{blue!15}69.73 & \underline{60.76} & \underline{83.00}
           & \cellcolor{blue!15}74.68 & \cellcolor{blue!15}92.86 & \underline{88.40} & \underline{98.30} & \underline{66.88} \\
       \addlinespace[3pt]\cmidrule(lr){3-12}
       &  & FineLIP DCI FT
           & 40.95 & 65.70 & 58.80 & 81.94
           & 36.30 & 60.22 & 52.36 & 76.56 & 47.10 \\
       &  & GOAL DCI FT
           & \underline{43.90} & \underline{68.60} & \underline{61.12} & \underline{83.30}
           & \underline{72.88} & \underline{91.68} & \underline{88.10} & \underline{98.10} & \underline{66.50} \\
       &  & \textbf{MulCLIP DCI FT}
           & \cellcolor{blue!15}{44.25} & \cellcolor{blue!15}{69.20} & \cellcolor{blue!15}{62.86} & \cellcolor{blue!15}{83.44}
           & \cellcolor{blue!15}{74.04} & \cellcolor{blue!15}{92.44} & \cellcolor{blue!15}{89.60} & \cellcolor{blue!15}{98.50} & \cellcolor{blue!15}67.69 \\
      \bottomrule
    \end{tabular}%
  }
  \vspace{-1.0em}
\end{table*}

\paragraph{In-domain Long Caption Retrieval.}  On Tab.~\ref{tab:cross_dci_docci}, MulCLIP establishes clear in-domain advantages on both DOCCI and DCI.   On DOCCI, MulCLIP achieves the highest R@1 among the compared adaptation methods across both evaluated backbones and both retrieval directions, improving the average of T2I and I2T R@1 over GOAL by 1.78 and 2.30 percentage points for ViT-B/16 and ViT-L/14, respectively, and outperforming FineLIP in both directions. On DCI, where description quality control is weaker than in DOCCI \citep{refDOCCI}, although GOAL benefits from its segment-filtering procedure, MulCLIP is able to achieve competitive performance with GOAL and continues to surpass FineLIP by a large margin.

\paragraph{Zero-shot Long-Caption Cross-Modal Retrieval.}
MulCLIP generalizes robustly across fine-grained long-text retrieval domains, as shown in Table~\ref{tab:cross_dci_docci}. When fine-tuned on DOCCI and tested on DCI, it surpasses GOAL on T2I R@1 by 4.95 percentage points with ViT-B/16 (69.08\% vs.\ 64.13\%) and 4.00 points with ViT-L/14 (72.93\% vs.\ 68.93\%). The I2T gains are 1.25 points with ViT-B/16 (67.13\% vs.\ 65.88\%) and 3.60 points with ViT-L/14 (72.03\% vs.\ 68.43\%). In the reverse setting (fine-tuned on DCI, evaluated on DOCCI), MulCLIP remains competitive with GOAL. On Urban1K, MulCLIP improves over GOAL on most recall metrics across both backbones and fine-tuning datasets.
\paragraph{Zero-shot Short Caption Retrieval.}
After fine-tuning on long-caption data, MulCLIP still performs strongly on short-caption benchmarks. In many cases it improves over the pretrained CLIP baseline and tends to be stronger on T2I while staying competitive on I2T. For example, on Flickr30k, with ViT-L/14 trained on DCI, MulCLIP reaches I2T R@1 of 89.6\% (vs.\ GOAL 88.1\%, CLIP 86.7\%); with ViT-B/16 trained on DOCCI, it attains T2I R@1 of 67.44\% (vs.\ 66.92\%, 63.20\%). On COCO, it continues to lead T2I R@1 when trained on DCI for both backbones, and otherwise stays within roughly 1--2\% of GOAL. For I2T, results are comparable, occasionally trailing GOAL by about 1--2\%. Additional evaluations across vision-language architectures and pretraining regimes are reported in the supplementary material Sec~\ref{sec:additional_backbones}.

\begin{table*}[t]
  \centering
  \caption{\textbf{Ablation study on different designs of MulCLIP on long (Urban, DCI, DOCCI) and short text (COCO, Flickr) dataset.} Overall, MulCLIP shows consistent improvements in long text understanding while remaining competitive at short text across baselines.}
  \label{tab:ablation_combined_b16}
  \resizebox{\linewidth}{!}{%
    \begin{tabular}{@{} l
                    c c c c c c c
                    c c c c c
                    @{}}
      \toprule
      \multirow{3}{*}{\textbf{Method}}
        & \multicolumn{7}{c}{\textbf{Long caption understanding}}
        & \multicolumn{5}{c}{\textbf{Short caption understanding}} \\
      \cmidrule(lr){2-8}\cmidrule(lr){9-13}
        & \multicolumn{2}{c}{\textbf{Urban-1k}}
        & \multicolumn{2}{c}{\textbf{DCI}}
        & \multicolumn{2}{c}{\textbf{DOCCI}}
        & \textbf{Avg (long)}
        & \multicolumn{2}{c}{\textbf{COCO}}
        & \multicolumn{2}{c}{\textbf{Flickr}}
        & \textbf{Avg (short)} \\
      \cmidrule(lr){2-3}\cmidrule(lr){4-5}\cmidrule(lr){6-7}\cmidrule(lr){9-10}\cmidrule(lr){11-12}
        & \textbf{T2I} & \textbf{I2T}
        & \textbf{T2I} & \textbf{I2T}
        & \textbf{T2I} & \textbf{I2T}
        &
        & \textbf{T2I} & \textbf{I2T}
        & \textbf{T2I} & \textbf{I2T}
        & \\
      \midrule

      Zeroshot CLIP
        & -- & --
        & -- & --
        & -- & --
        & --
        & 33.95 & 54.14
        & 63.20 & 82.90
        & 58.55 \\
      \addlinespace[1pt] \cmidrule(lr){1-13}

      Global only
        & 71.2 & 80.9
        & 65.0 & 63.9
        & 81.1 & 79.9
        & 73.67
        & \cellcolor{blue!15}38.03 & 54.98
        & 66.80 & \cellcolor{blue!15}84.00
        & \underline{60.95} \\
      W/o SAP
        & 74.4 & 80.1
        & 65.4 & 64.1
        & 80.6 & 78.9
        & 73.92
        & \underline{38.02} & \cellcolor{blue!15}55.74
        & \cellcolor{blue!15}67.68 & \underline{83.90}
        & \cellcolor{blue!15}{61.34} \\
      W/o WPR
        & 73.1 & 80.0
        & 66.4 & \underline{65.6}
        & \cellcolor{blue!15}82.9 & \cellcolor{blue!15}81.6
        & 74.93
        & 37.12 & 54.88
        & 65.74 & 83.00
        & 60.19 \\
      \textbf{MulCLIP (ours)}
        & \cellcolor{blue!15}77.3 & \cellcolor{blue!15}84.0
        & \cellcolor{blue!15}69.1 & \cellcolor{blue!15}67.1
        & 82.2 & 80.3
        & \cellcolor{blue!15}76.67
        & 37.68 & 54.76
        & \underline{67.44} & 81.90
        & 60.45 \\
      \addlinespace[1pt] \cmidrule(lr){1-13}
      MulCLIP w CLIM
        & 68.3 & 78.8
        & 64.0 & 62.8
        & 78.5 & 77.2
        & 71.60
        & 34.77 & 48.22
        & 64.44 & 78.50
        & 56.48 \\
      \bottomrule
    \end{tabular}%
  }
  \vspace{-0.3em}
\end{table*}

\begin{figure*}[t]
    \centering
    \makebox[\textwidth][c]{%
    \hspace*{3mm}%
    \includegraphics[
        height=0.38\textheight,
        keepaspectratio
    ]{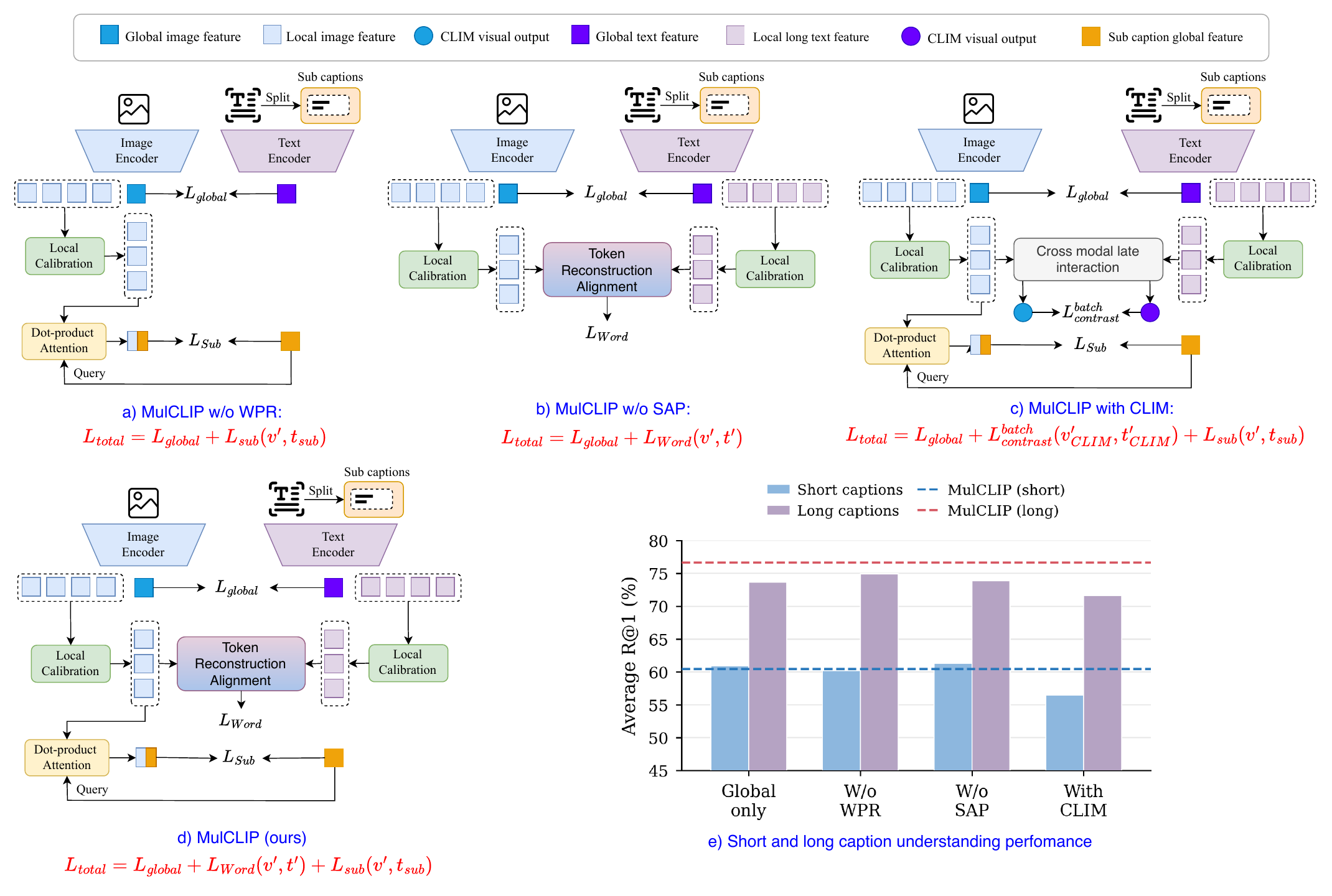}%
    \hspace*{-3mm}%
}
    \caption{\textbf{Visualization of ablation study designs and their formulations.} Starting from the global-only setting $L_{\text{global}}$, adding WPR and SAP improves average long-caption retrieval performance. Replacing the token reconstruction objective with CLIM yields inferior performance, suggesting incompatibility with our subcaption grounding design.}
    \label{fig:core_ablation}
\end{figure*}

\section{Ablation Study \& Analysis}
\label{subsec:abla}
We conduct extensive ablation studies to evaluate the contribution of each component in MulCLIP, using ViT B/16 checkpoints fine-tuned on DOCCI and tested on long/short image--text retrieval. We further report the degradation in zero-shot classification  and quantitative localization fidelity in the supplementary material Sec. \ref{app:classification} and Sec.~\ref{app:fg_quantitative_result}.

To highlight the role of each component in MulCLIP, we consider the variants summarized in Fig.~\ref{fig:core_ablation}. To further examine whether batch-wise local token alignment conflicts with subcaption-level supervision in alternative design, we evaluate \textbf{``MulCLIP w CLIM'',} which keeps the full MulCLIP objective but replaces the word-patch reconstruction mechanism (within-sample contrastive) with the Cross-modal Late Interaction Module (CLIM) mechanism (batch-wise contrastive) ~\citep{refFineLIP}. Other ablations are reported in the supplementary material Sec.~\ref{tab:addition_ablation}.
\vspace{-0.5em}
\paragraph{Core Components Contribution.}

Tab.~\ref{tab:ablation_combined_b16} isolates the role of each MulCLIP component on ViT-B/16 across long- and short-caption retrieval. Adding calibrated token-level word-patch reconstruction "W/o SAP" consistently strengthens both long-caption benchmarks (Urban1k/DCI/DOCCI) and short-caption retrieval (COCO/Flickr) over the "Global only" baseline, indicating that calibration + WPR provides a robust fine-grained alignment signal. Introducing SAP on top of this yields the best overall long-caption performance (highest long-caption average), showing that subcaption-to-region grounding is critical for structured long-text understanding.

\subsection{Objective Compatibility in the Local Token Space.}
We examine the compatibility of token-level alignment with subcaption-level supervision by comparing two settings in the decoupled local token space: "MulCLIP w/o WPR" (global+SAP) and "MulCLIP w CLIM" (global+SAP+CLIM), as illustrated in Fig.~\ref{fig:core_ablation}. From Tab.~\ref{tab:ablation_combined_b16}, we observe a clear gap between these two settings: introducing CLIM on top of SAP leads to a substantial average drop of about 4-5\%, rather than an improvement. This suggests that CLIM may not directly integrate with subcaption-level supervision, as the resulting local token learning signal can interfere with SAP's sentence-to-region grounding. In contrast, MulCLIP achieves stronger overall results, indicating that our framework combines global alignment and subcaption grounding in a more compatible manner, and is effective for unified long-context training.

\label{sec:analysis}
\begin{figure*}[t]
    \centering  \includegraphics[width=0.95\textwidth]
    {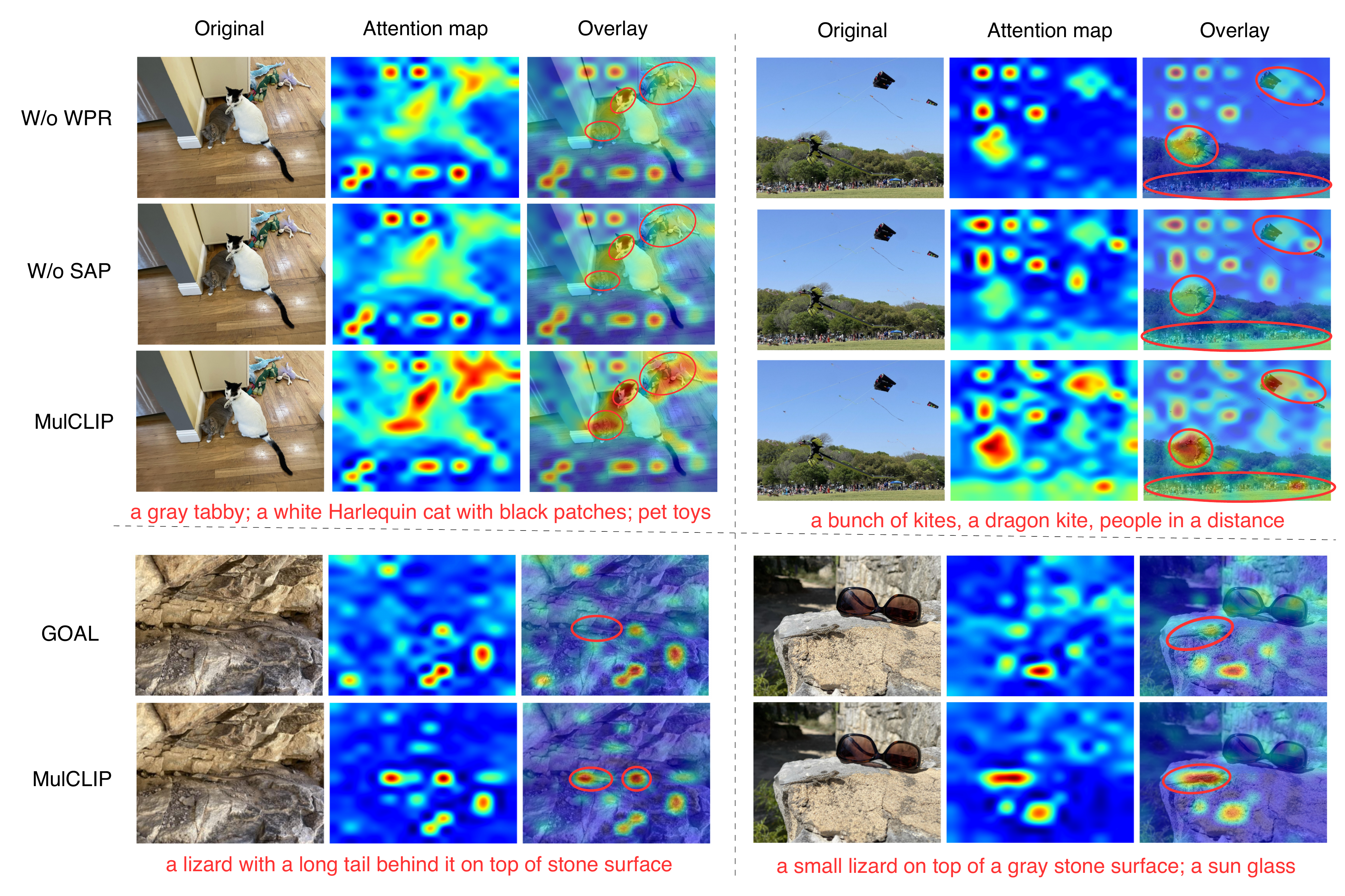}
     \caption{\textbf{Qualitative comparison of attention maps.}
From left to right, we show: (1) the original image, (2) the attention heatmap, and (3) the overlay of the heatmap on the image. Across diverse scenes, MulCLIP produces more concentrated attention patterns that visually correspond better to the queried fine-grained details. Red circles highlight regions where MulCLIP demonstrates effective attention localization.}
    \label{fig:qual_map}
    \vspace{-0.2cm}
\end{figure*}

\subsection{Fine-grained qualitative results}

Fig.~\ref{fig:qual_map} compares ViT-B/16 attention maps of \textbf{GOAL}, the ablations (\textbf{W/o SAP}, \textbf{W/o WPR}), and our full \textbf{MulCLIP} model. MulCLIP consistently captures local details more precisely than any of the baselines. Both MulCLIP and its ablations can detect subtle cues such as the camouflaged long-tailed lizard on the rocks. However, while the ablations attend to these details, MulCLIP produces sharper and more semantically aligned activations; in contrast, "W/o SAP" and GOAL yield more diffuse responses, whereas "W/o WPR" produces less diffuse but more fragmented patterns that often miss broader contextual regions (e.g., eyeglasses and distant pedestrians). Notably, in this example, GOAL assigns little attention to the camouflaged lizard, despite its use of SAM-based region proposals to support localization, revealing a blind spot compared to MulCLIP's self-learned alignment mechanism. These qualitative comparisons reinforce the quantitative results, indicating that MulCLIP effectively balances global comprehension with fine-grained localization while avoiding the drawbacks of external region-proposal modules.

\vspace{-5pt}
\section{Conclusion and Future Work}
\label{sec:conclusion}
We presented MulCLIP, a proposal-free adaptation framework that brings multi-level alignment to CLIP-style models. Experiments demonstrate improved long-caption retrieval while maintaining competitive short-caption performance. Ablation studies support the complementary roles of global, subcaption-level, and word-level alignment.

Robustness to noisy or less structured captions remains to be established. Extending MulCLIP to learn reliable correspondences from such data could broaden its applicability beyond densely annotated datasets and support its use in large-scale vision-language pretraining, providing a promising direction for future work.

\section*{Acknowledgments and Disclosure of Funding}

This work received computational support from Vinmotion. The authors declare no additional competing interests.

\bibliographystyle{plainnat}
\bibliography{references}

\begin{thebibliography}{32}
\providecommand{\natexlab}[1]{#1}
\providecommand{\url}[1]{\texttt{#1}}
\expandafter\ifx\csname urlstyle\endcsname\relax
  \providecommand{\doi}[1]{doi: #1}\else
  \providecommand{\doi}{doi: \begingroup \urlstyle{rm}\Url}\fi

\bibitem[Asokan et~al.(2025)Asokan, Wu, and Albreiki]{refFineLIP}
Mothilal Asokan, Kebin Wu, and Fatima Albreiki.
\newblock Finelip: Extending clip's reach via fine-grained alignment with
  longer text inputs.
\newblock In \emph{Proceedings of the IEEE/CVF Conference on Computer Vision
  and Pattern Recognition (CVPR)}, pages 14495--14504, June 2025.

\bibitem[Bianchi et~al.(2024)Bianchi, Carrara, Messina, Gennaro, and
  Falchi]{citeFG-OVD}
Lorenzo Bianchi, Fabio Carrara, Nicola Messina, Claudio Gennaro, and Fabrizio
  Falchi.
\newblock The devil is in the fine-grained details: Evaluating open-vocabulary
  object detectors for fine-grained understanding, 2024.
\newblock URL \url{https://arxiv.org/abs/2311.17518}.

\bibitem[Bica et~al.(2024)Bica, Ilić, Bauer, Erdogan, Bošnjak, Kaplanis,
  Gritsenko, Minderer, Blundell, Pascanu, and Mitrović]{refSPARC}
Ioana Bica, Anastasija Ilić, Matthias Bauer, Goker Erdogan, Matko Bošnjak,
  Christos Kaplanis, Alexey~A. Gritsenko, Matthias Minderer, Charles Blundell,
  Razvan Pascanu, and Jovana Mitrović.
\newblock Improving fine-grained understanding in image-text pre-training,
  2024.
\newblock URL \url{https://arxiv.org/abs/2401.09865}.

\bibitem[Bolya et~al.(2023)Bolya, Fu, Dai, Zhang, Feichtenhofer, and
  Hoffman]{refTOME}
Daniel Bolya, Cheng-Yang Fu, Xiaoliang Dai, Peizhao Zhang, Christoph
  Feichtenhofer, and Judy Hoffman.
\newblock Token merging: Your vit but faster.
\newblock In \emph{The Eleventh International Conference on Learning
  Representations}, 2023.
\newblock URL \url{https://openreview.net/forum?id=JroZRaRw7Eu}.

\bibitem[Choi et~al.(2025)Choi, Jang, and Eom]{refGOAL}
Hyungyu Choi, Young~Kyun Jang, and Chanho Eom.
\newblock Goal: Global-local object alignment learning.
\newblock \emph{ArXiv}, abs/2503.17782, 2025.
\newblock URL \url{https://api.semanticscholar.org/CorpusID:277271667}.

\bibitem[Fu et~al.(2024)Fu, Zhang, Xia, and Mao]{refLAPS}
Zheren Fu, Lei Zhang, Hou Xia, and Zhendong Mao.
\newblock Linguistic-aware patch slimming framework for fine-grained
  cross-modal alignment.
\newblock In \emph{Proceedings of the IEEE/CVF Conference on Computer Vision
  and Pattern Recognition (CVPR)}, pages 26307--26316, June 2024.

\bibitem[Han et~al.(2023)Han, Zhu, Lao, and Jiang]{refZeroshotREC}
Zeyu Han, Fangrui Zhu, Qianru Lao, and Huaizu Jiang.
\newblock Zero-shot referring expression comprehension via structural
  similarity between images and captions.
\newblock \emph{2024 IEEE/CVF Conference on Computer Vision and Pattern
  Recognition (CVPR)}, pages 14364--14375, 2023.
\newblock URL \url{https://api.semanticscholar.org/CorpusID:265466308}.

\bibitem[Hendrycks and Gimpel(2016)]{gelu}
Dan Hendrycks and Kevin Gimpel.
\newblock Gaussian error linear units (gelus).
\newblock \emph{arXiv preprint arXiv:1606.08415}, 2016.

\bibitem[Huang et~al.(2021)Huang, Shen, Lungren, and Yeung]{refGLoRIA}
Shih-Cheng Huang, Liyue Shen, Matthew~P. Lungren, and Serena Yeung.
\newblock Gloria: A multimodal global-local representation learning framework
  for label-efficient medical image recognition.
\newblock In \emph{2021 IEEE/CVF International Conference on Computer Vision
  (ICCV)}, pages 3922--3931, 2021.
\newblock \doi{10.1109/ICCV48922.2021.00391}.

\bibitem[Jia et~al.(2021)Jia, Yang, Xia, Chen, Parekh, Pham, Le, Sung, Li, and
  Duerig]{refALIGN}
Chao Jia, Yinfei Yang, Ye~Xia, Yi-Ting Chen, Zarana Parekh, Hieu Pham, Quoc Le,
  Yun-Hsuan Sung, Zhen Li, and Tom Duerig.
\newblock Scaling up visual and vision-language representation learning with
  noisy text supervision.
\newblock In \emph{International conference on machine learning}, pages
  4904--4916. PMLR, 2021.

\bibitem[Lai et~al.(2025)Lai, Saveris, Chen, Chen, Zhang, Zhang, Hu, Tebar,
  Gan, Grasch, Cao, and Yang]{refRevisit}
Zhengfeng Lai, Vasileios Saveris, Chen Chen, Hong-You Chen, Haotian Zhang,
  Bowen Zhang, Wenze Hu, Juan~Lao Tebar, Zhe Gan, Peter Grasch, Meng Cao, and
  Yinfei Yang.
\newblock Revisit large-scale image-caption data in pre-training multimodal
  foundation models.
\newblock In \emph{The Thirteenth International Conference on Learning
  Representations}, 2025.
\newblock URL \url{https://openreview.net/forum?id=TWnUgSAWNw}.

\bibitem[Li et~al.(2022)Li, Li, Xiong, and Hoi]{refBLIP}
Junnan Li, Dongxu Li, Caiming Xiong, and Steven Hoi.
\newblock Blip: Bootstrapping language-image pre-training for unified
  vision-language understanding and generation.
\newblock In \emph{International conference on machine learning}, pages
  12888--12900. PMLR, 2022.

\bibitem[Li et~al.(2023)Li, Li, Savarese, and Hoi]{refBLIP2}
Junnan Li, Dongxu Li, Silvio Savarese, and Steven Hoi.
\newblock {BLIP}-2: Bootstrapping language-image pre-training with frozen image
  encoders and large language models.
\newblock In \emph{Proceedings of the 40th International Conference on Machine
  Learning}, volume 202 of \emph{Proceedings of Machine Learning Research},
  pages 19730--19742. PMLR, 2023.
\newblock URL \url{https://proceedings.mlr.press/v202/li23q.html}.

\bibitem[Lin et~al.(2015)Lin, Maire, Belongie, Bourdev, Girshick, Hays, Perona,
  Ramanan, Zitnick, and Dollár]{refCOCO}
Tsung-Yi Lin, Michael Maire, Serge Belongie, Lubomir Bourdev, Ross Girshick,
  James Hays, Pietro Perona, Deva Ramanan, C.~Lawrence Zitnick, and Piotr
  Dollár.
\newblock Microsoft coco: Common objects in context, 2015.
\newblock URL \url{https://arxiv.org/abs/1405.0312}.

\bibitem[Liu et~al.(2023)Liu, Li, Wu, and Lee]{refLLAVA}
Haotian Liu, Chunyuan Li, Qingyang Wu, and Yong~Jae Lee.
\newblock Visual instruction tuning.
\newblock In \emph{Thirty-seventh Conference on Neural Information Processing
  Systems}, 2023.
\newblock URL \url{https://openreview.net/forum?id=w0H2xGHlkw}.

\bibitem[Najdenkoska et~al.(2025)Najdenkoska, Derakhshani, Asano, Noord,
  Worring, and Snoek]{refTulip}
Ivona Najdenkoska, Mohammad~Mahdi Derakhshani, Yuki~M Asano, Nanne~Van Noord,
  Marcel Worring, and Cees G.~M. Snoek.
\newblock {TULIP}: Token-length upgraded {CLIP}.
\newblock In \emph{The Thirteenth International Conference on Learning
  Representations}, 2025.
\newblock URL \url{https://openreview.net/forum?id=r9oqHOdoHf}.

\bibitem[Nichol et~al.(2022)Nichol, Dhariwal, Ramesh, Shyam, Mishkin, McGrew,
  Sutskever, and Chen]{refGLIDE}
Alex Nichol, Prafulla Dhariwal, Aditya Ramesh, Pranav Shyam, Pamela Mishkin,
  Bob McGrew, Ilya Sutskever, and Mark Chen.
\newblock Glide: Towards photorealistic image generation and editing with
  text-guided diffusion models, 2022.
\newblock URL \url{https://arxiv.org/abs/2112.10741}.

\bibitem[Onoe et~al.(2024)Onoe, Rane, Berger, Bitton, Cho, Garg, Ku, Parekh,
  Pont-Tuset, Tanzer, Wang, and Baldridge]{refDOCCI}
Yasumasa Onoe, Sunayana Rane, Zachary Berger, Yonatan Bitton, Jaemin Cho,
  Roopal Garg, Alexander Ku, Zarana Parekh, Jordi Pont-Tuset, Garrett Tanzer,
  Su~Wang, and Jason Baldridge.
\newblock Docci: Descriptions of connected and contrasting images.
\newblock \emph{ArXiv}, abs/2404.19753, 2024.
\newblock URL \url{https://api.semanticscholar.org/CorpusID:269457405}.

\bibitem[Plummer et~al.(2016)Plummer, Wang, Cervantes, Caicedo, Hockenmaier,
  and Lazebnik]{refFlick30k}
Bryan~A. Plummer, Liwei Wang, Chris~M. Cervantes, Juan~C. Caicedo, Julia
  Hockenmaier, and Svetlana Lazebnik.
\newblock Flickr30k entities: Collecting region-to-phrase correspondences for
  richer image-to-sentence models, 2016.
\newblock URL \url{https://arxiv.org/abs/1505.04870}.

\bibitem[Radford et~al.(2021)Radford, Kim, Hallacy, Ramesh, Goh, Agarwal,
  Sastry, Askell, Mishkin, Clark, Krueger, and Sutskever]{refCLIP}
Alec Radford, Jong~Wook Kim, Chris Hallacy, Aditya Ramesh, Gabriel Goh,
  Sandhini Agarwal, Girish Sastry, Amanda Askell, Pamela Mishkin, Jack Clark,
  Gretchen Krueger, and Ilya Sutskever.
\newblock Learning transferable visual models from natural language
  supervision.
\newblock In \emph{International Conference on Machine Learning}, 2021.
\newblock URL \url{https://api.semanticscholar.org/CorpusID:231591445}.

\bibitem[Rombach et~al.(2022)Rombach, Blattmann, Lorenz, Esser, and
  Ommer]{refStableDiffusion}
Robin Rombach, Andreas Blattmann, Dominik Lorenz, Patrick Esser, and Björn
  Ommer.
\newblock High-resolution image synthesis with latent diffusion models, 2022.
\newblock URL \url{https://arxiv.org/abs/2112.10752}.

\bibitem[Tong et~al.(2024)Tong, Liu, Zhai, Ma, LeCun, and Xie]{refEyewideshut}
Shengbang Tong, Zhuang Liu, Yuexiang Zhai, Yi~Ma, Yann LeCun, and Saining Xie.
\newblock Eyes wide shut? exploring the visual shortcomings of multimodal llms,
  2024.
\newblock URL \url{https://arxiv.org/abs/2401.06209}.

\bibitem[Urbanek et~al.(2024)Urbanek, Bordes, Astolfi, Williamson, Sharma, and
  Romero-Soriano]{refDCI}
Jack Urbanek, Florian Bordes, Pietro Astolfi, Mary Williamson, Vasu Sharma, and
  Adriana Romero-Soriano.
\newblock A picture is worth more than 77 text tokens: Evaluating clip-style
  models on dense captions, 2024.
\newblock URL \url{https://arxiv.org/abs/2312.08578}.

\bibitem[Wu et~al.(2024)Wu, Zheng, Ma, Lu, Guo, Zhang, Chen, Guo, Shen, and
  Zha]{refLotLIP}
Wei Wu, Kecheng Zheng, Shuailei Ma, Fan Lu, Yuxin Guo, Yifei Zhang, Wei Chen,
  Qingpei Guo, Yujun Shen, and Zheng-Jun Zha.
\newblock Lotlip: Improving language-image pre-training for long text
  understanding, 2024.
\newblock URL \url{https://arxiv.org/abs/2410.05249}.

\bibitem[Xiao et~al.(2025)Xiao, Kim, Georgescu, Akata, and Alaniz]{refFLAIR}
Rui Xiao, Sanghwan Kim, Mariana-Iuliana Georgescu, Zeynep Akata, and Stephan
  Alaniz.
\newblock Flair: Vlm with fine-grained language-informed image representations.
\newblock In \emph{Proceedings of the Computer Vision and Pattern Recognition
  Conference (CVPR)}, pages 24884--24894, June 2025.

\bibitem[Xie et~al.(2025)Xie, Wang, Kong, Li, Liang, Zhang, Leng, and
  Yin]{refFG_CLIP}
Chunyu Xie, Bin Wang, Fanjing Kong, Jincheng Li, Dawei Liang, Gengshen Zhang,
  Dawei Leng, and Yuhui Yin.
\newblock {FG}-{CLIP}: Fine-grained visual and textual alignment.
\newblock In \emph{Forty-second International Conference on Machine Learning},
  2025.
\newblock URL \url{https://openreview.net/forum?id=eih5Gy3Pjt}.

\bibitem[Yao et~al.(2021)Yao, Huang, Hou, Lu, Niu, Xu, Liang, Li, Jiang, and
  Xu]{refFILIP}
Lewei Yao, Runhui Huang, Lu~Hou, Guansong Lu, Minzhe Niu, Hang Xu, Xiaodan
  Liang, Zhenguo Li, Xin Jiang, and Chunjing Xu.
\newblock Filip: Fine-grained interactive language-image pre-training, 2021.
\newblock URL \url{https://arxiv.org/abs/2111.07783}.

\bibitem[Zhai et~al.(2023)Zhai, Mustafa, Kolesnikov, and Beyer]{refSiglip}
Xiaohua Zhai, Basil Mustafa, Alexander Kolesnikov, and Lucas Beyer.
\newblock Sigmoid loss for language image pre-training.
\newblock In \emph{Proceedings of the IEEE/CVF international conference on
  computer vision}, pages 11975--11986, 2023.

\bibitem[Zhang et~al.(2024)Zhang, Zhang, wen Dong, Zang, and Wang]{refLongCLIP}
Beichen Zhang, Pan Zhang, Xiao wen Dong, Yuhang Zang, and Jiaqi Wang.
\newblock Long-clip: Unlocking the long-text capability of clip.
\newblock In \emph{European Conference on Computer Vision}, 2024.
\newblock URL \url{https://api.semanticscholar.org/CorpusID:268667201}.

\bibitem[Zhang et~al.(2025)Zhang, Xu, Usuyama, Xu, Bagga, Tinn, Preston, Rao,
  Wei, Valluri, Wong, Tupini, Wang, Mazzola, Shukla, Liden, Gao, Crabtree,
  Piening, Bifulco, Lungren, Naumann, Wang, and Poon]{refBiomedCLIP}
Sheng Zhang, Yanbo Xu, Naoto Usuyama, Hanwen Xu, Jaspreet Bagga, Robert Tinn,
  Sam Preston, Rajesh Rao, Mu~Wei, Naveen Valluri, Cliff Wong, Andrea Tupini,
  Yu~Wang, Matt Mazzola, Swadheen Shukla, Lars Liden, Jianfeng Gao, Angela
  Crabtree, Brian Piening, Carlo Bifulco, Matthew~P. Lungren, Tristan Naumann,
  Sheng Wang, and Hoifung Poon.
\newblock Biomedclip: a multimodal biomedical foundation model pretrained from
  fifteen million scientific image-text pairs, 2025.
\newblock URL \url{https://arxiv.org/abs/2303.00915}.

\bibitem[Zheng et~al.(2024)Zheng, Zhang, Wu, Lu, Ma, Jin, Chen, and
  Shen]{refDreamLIP}
Kecheng Zheng, Yifei Zhang, Wei Wu, Fan Lu, Shuailei Ma, Xin Jin, Wei Chen, and
  Yujun Shen.
\newblock Dreamlip: Language-image pre-training with long captions.
\newblock \emph{ArXiv}, abs/2403.17007, 2024.
\newblock URL \url{https://api.semanticscholar.org/CorpusID:268681239}.

\bibitem[Zong et~al.(2022)Zong, Li, Song, Wang, Qiao, Leng, and
  Liu]{refSlim_VIT}
Zhuofan Zong, Kunchang Li, Guanglu Song, Yali Wang, Yu~Qiao, Biao Leng, and
  Yu~Liu.
\newblock Self-slimmed vision transformer, 2022.
\newblock URL \url{https://arxiv.org/abs/2111.12624}.

\end{thebibliography}

\newpage
\appendix
\onecolumn
\label{sec:appendix}

\section{Additional Implementation Details}
\label{app:implementation}

MulCLIP is implemented on top of a CLIP-style dual encoder loaded through the LongCLIP interface. We use ViT-B/16 and ViT-L/14 as visual backbones. For each training sample, we use an image, a long caption, and a short summary caption. Long captions are split into sentence-level detail captions using period-based segmentation. Since the first sentence often functions as a summary, detail-caption construction starts from the remaining sentences. Adjacent detail sentences are randomly merged with probability 0.5. Very short or invalid fragments are removed. If no valid detail caption is available, the short summary caption is used as a fallback. We sample at most \(M_{\max}=15\) detail captions per image. The main optimization and training hyperparameters are summarized in Table~\ref{tab:implementation_details}.

\begin{table}[htbp]
\centering
\small
\begin{tabular}{ll}
\toprule
\textbf{Setting} & \textbf{Value} \\
\midrule
Backbones & Pretrained ViT-B/16, ViT-L/14 \\
Text context length & 248 \\
Optimizer & AdamW \\
Epochs & 8 \\
Backbone learning rate & \(1\times10^{-5}\) \\
Calibration-module learning rate & \(2\times10^{-4}\) \\
Weight decay & 0.05 \\
Scheduler & Cosine decay \\
Warmup & 200 iterations \\
Batch size & 16 for both ViT-B/16 and ViT-L/14 \\
Maximum detail captions & 15 \\
Loss weights & \(1.0, 0.5, 1.0, 1.0\) \\
Mixed precision & Enabled \\
Random seed & 42 \\
Hardware & NVIDIA A5000 for ViT-B/16 and NVIDIA A100 for ViT-L/14 \\
\bottomrule
\end{tabular}
\caption{Implementation and training hyperparameters used for MulCLIP fine-tuning.}
\label{tab:implementation_details}
\vspace{-0.8em}
\end{table}

\begin{table}[htbp]
  \centering
  \captionsetup{font=scriptsize,skip=2pt}
  \caption{\textbf{Classification performance on CIFAR and ImageNet variants.}}
  \label{tab:docci_ft_classification_only}
  {\scriptsize
  \setlength{\tabcolsep}{2.5pt}
  \renewcommand{\arraystretch}{0.88}
  \resizebox{\linewidth}{!}{%
    \begin{tabular}{@{} l l c c c c c @{}}
      \toprule
      & \multirow{2}{*}{\textbf{Method}}
      & \multicolumn{5}{c}{\textbf{Top-1 Accuracy (\%)}} \\
      \cmidrule(lr){3-7}
      &  & \textbf{CIFAR-10} & \textbf{CIFAR-100} & \textbf{ImageNet-O} & \textbf{ImageNet-V2} & \textbf{Avg} \\
      \midrule
      \multirow{4}{*}{\rotatebox{90}{\scriptsize ViT-B/16}}
        & GOAL
          & 84.95 & \underline{55.41} & 42.15 & 49.85 & 58.09 \\
        & W/o SAP
          & 81.36 & 52.89 & \underline{43.00} & 50.94 & 57.05 \\
        & W/o WPR
          & \underline{84.98} & 55.07 & 41.65 & \underline{51.82} & \underline{58.38} \\
        & MulCLIP
          & \cellcolor{blue!15}86.33 & \cellcolor{blue!15}60.34 & \cellcolor{blue!15}43.80 & \cellcolor{blue!15}52.13 & \cellcolor{blue!15}60.65 \\
      \bottomrule
    \end{tabular}%
  }}
\end{table}

\section{Zero-shot classification performance.}
\label{app:classification}
Tab.~\ref{tab:docci_ft_classification_only} reports zero-shot top-1 accuracy for DOCCI fine-tuned checkpoints. With ViT-B/16, MulCLIP is best on all four datasets and achieves the highest average accuracy (60.65\%), outperforming GOAL (58.09\%) by 2.6\%, with clear gains on CIFAR-100 (60.34\% vs.\ 55.41\%) and ImageNet-V2 (52.13\% vs.\ 49.85\%); the ablations consistently underperform, indicating that both SAP and WPR help preserve classification after long-caption fine-tuning.

\section{Adaptation Across Pretraining Regimes}
\label{sec:additional_backbones}

We further evaluate MulCLIP on BLIP-2-Giant \citep{refBLIP2} and FG-CLIP ViT-B/16 to test whether the improvements observed with CLIP carry over to other pretrained models. BLIP-2 extends the evaluation to a different vision-language architecture, while FG-CLIP addresses a complementary question: whether MulCLIP remains beneficial when the backbone has already been pretrained on long descriptions.

\subsection{Adaptation from a Short-Text-Pretrained Checkpoint}
\label{sec:blip2_adaptation}
Following GOAL's evaluation setup, we fine-tune
BLIP-2-Giant separately on DOCCI and DCI using the
respective splits from GOAL's released code.
We compare CLIP-style global contrastive fine-tuning,
GOAL, and MulCLIP on in-domain text-to-image and
image-to-text retrieval. As shown in Table~\ref{tab:blip2_adaptation}, MulCLIP outperforms both baselines across all reported metrics on DOCCI and DCI.
\begin{table*}[ht!]
  \centering
  \caption{
  \textbf{Retrieval results on BLIP-2-Giant.}
  MulCLIP outperforms both baselines on DOCCI and DCI
  across all reported metrics (Recall@k, \%).
  Blue shading and underlining indicate the best and
  second-best results.
}
  \label{tab:blip2_adaptation}
  \resizebox{0.75\linewidth}{!}{%
    \begin{tabular}{@{} l l c c c c @{}}
      \toprule
      \multirow{2}{*}{\textbf{Dataset}}
      & \multirow{2}{*}{\textbf{Method}}
      & \multicolumn{2}{c}{\textbf{Text-to-Image}}
      & \multicolumn{2}{c}{\textbf{Image-to-Text}} \\
      \cmidrule(lr){3-4}\cmidrule(lr){5-6}
      & & R@1 & R@5 & R@1 & R@5 \\
      \midrule

      \multirow{3}{*}{DOCCI}
      & BLIP-2 + CLIP
        & 23.45 & 54.96 & 26.16 & 57.53 \\

      & BLIP-2 + GOAL
        & \underline{64.63} & \underline{90.02}
        & \underline{61.86} & \underline{88.47} \\

      & \textbf{BLIP-2 + MulCLIP (ours)}
        & \cellcolor{blue!15}{77.55}
        & \cellcolor{blue!15}{95.65}
        & \cellcolor{blue!15}{75.75}
        & \cellcolor{blue!15}{95.37} \\

      \midrule

      \multirow{3}{*}{DCI}
      & BLIP-2 + CLIP
        & 22.81 & 52.33 & 20.11 & 50.28 \\

      & BLIP-2 + GOAL
        & \underline{50.88} & \underline{77.49}
        & \underline{50.38} & \underline{77.49} \\

      & \textbf{BLIP-2 + MulCLIP (ours)}
        & \cellcolor{blue!15}{76.64}
        & \cellcolor{blue!15}{91.65}
        & \cellcolor{blue!15}{73.64}
        & \cellcolor{blue!15}{91.40} \\

      \bottomrule
    \end{tabular}%
  }
\end{table*}

\subsection{Adaptation from a Long-Text-Pretrained Checkpoint}
\label{sec:fgclip_adaptation}

We use FG-CLIP ViT-B/16 to examine whether MulCLIP provides additional benefits when the initial checkpoint already incorporates large-scale long-text pretraining. Starting from the same checkpoint, we compare global contrastive fine-tuning, GOAL, and MulCLIP on DCI. We evaluate both target-domain retrieval and transfer to four non-target benchmarks, including the original zero-shot checkpoint as a reference. As shown in Table~\ref{tab:fgclip_adaptation}, MulCLIP performs best among the fine-tuned methods, although it remains below the zero-shot checkpoint on the non-target benchmarks.

\begin{table*}[ht!]
  \centering
  \caption{
  \textbf{DCI adaptation of FG-CLIP ViT-B/16.}
  MulCLIP achieves the highest R@1 (\%) among fine-tuned
  methods across all benchmarks.
  Blue shading and underlining indicate the best and
  second-best fine-tuning results; zero-shot is a reference.
}
  \label{tab:fgclip_adaptation}
  \resizebox{0.95\linewidth}{!}{%
    \begin{tabular}{@{} l
                     c c
                     c c
                     c c
                     c c
                     c c @{}}
      \toprule
      \multirow{2}{*}{\textbf{Method}}
      & \multicolumn{2}{c}{\textbf{DCI}}
      & \multicolumn{2}{c}{\textbf{DOCCI}}
      & \multicolumn{2}{c}{\textbf{Urban1K}}
      & \multicolumn{2}{c}{\textbf{Flickr30K}}
      & \multicolumn{2}{c}{\textbf{COCO}} \\
      \cmidrule(lr){2-3}
      \cmidrule(lr){4-5}
      \cmidrule(lr){6-7}
      \cmidrule(lr){8-9}
      \cmidrule(lr){10-11}
      & T$\to$I & I$\to$T
      & T$\to$I & I$\to$T
      & T$\to$I & I$\to$T
      & T$\to$I & I$\to$T
      & T$\to$I & I$\to$T \\
      \midrule

      Zero-shot
        & 75.64 & 75.64
        & 84.65 & 84.14
        & 93.40 & 94.00
        & 75.48 & 90.90
        & 45.44 & 64.14 \\

      \midrule

      Global FT
        & 76.34 & 75.49
        & 78.76 & 77.33
        & 80.90 & 85.40
        & \underline{68.02} & 83.90
        & \underline{39.34} & 56.98 \\

      GOAL FT
        & \underline{79.69} & \underline{78.44}
        & \underline{80.39} & \underline{79.92}
        & \underline{85.50} & \underline{89.11}
        & 67.64 & \underline{87.29}
        & 38.77 & \underline{60.32} \\

      \textbf{MulCLIP FT}
        & \cellcolor{blue!15}{80.89}
        & \cellcolor{blue!15}{79.29}
        & \cellcolor{blue!15}{82.00}
        & \cellcolor{blue!15}{82.59}
        & \cellcolor{blue!15}{89.30}
        & \cellcolor{blue!15}{90.60}
        & \cellcolor{blue!15}{73.54}
        & \cellcolor{blue!15}{88.40}
        & \cellcolor{blue!15}{43.11}
        & \cellcolor{blue!15}{61.62} \\

      \bottomrule
    \end{tabular}%
  }
\end{table*}

\section{Computational Cost}
\label{sec:computational_cost}

\paragraph{Training cost.}
We compare GOAL, FineLIP, and MulCLIP on DOCCI using
ViT-B/16, batch size 16, and the same GPU.
GOAL is trained for 10 epochs, while FineLIP and MulCLIP
are trained for 8 epochs. We report both total training
time and time per epoch, together with parameter counts,
GPU-hours, memory usage, and energy consumption.

\begin{table*}[ht!]
  \centering
  \caption{
    \textbf{Training cost on DOCCI with ViT-B/16.}
    Blue shading marks the lowest measured time,
    memory usage, or energy consumption.
  }
  \label{tab:computational_cost}
  \resizebox{\linewidth}{!}{%
    \begin{tabular}{@{} l c c c c c c c c @{}}
      \toprule
      \textbf{Method}
      & \shortstack{\textbf{Params}\\(M)}
      & \shortstack{\textbf{Added}\\\textbf{params (M)}}
      & \textbf{Epochs}
      & \shortstack{\textbf{Train}\\(min)}
      & \shortstack{\textbf{Time}\\(min/epoch)}
      & \shortstack{\textbf{GPU}\\\textbf{hours}}
      & \shortstack{\textbf{Memory}\\(GiB)}
      & \shortstack{\textbf{Energy}\\(Wh)} \\
      \midrule

      GOAL
        & 149.708 & +0.000 & 10
        & 41.93 & 4.19 & 0.699
        & 15.79 & 100.79 \\

      FineLIP
        & 149.950 & +0.242 & 8
        & 28.42 & 3.55 & 0.474
        & \cellcolor{blue!15}{6.73}
        & 67.57 \\

      \textbf{MulCLIP}
        & 150.476 & +0.768 & 8
        & \cellcolor{blue!15}{23.87}
        & \cellcolor{blue!15}{2.98}
        & \cellcolor{blue!15}{0.398}
        & 12.77
        & \cellcolor{blue!15}{56.15} \\

      \bottomrule
    \end{tabular}%
  }
\end{table*}

Table~\ref{tab:computational_cost} shows that MulCLIP
has the lowest measured training time per epoch,
total training time, and energy consumption among
the compared methods. It introduces 0.768M additional
training parameters, while its memory usage lies
between FineLIP and GOAL.

\begin{table}[t]
    \centering
    \caption{\textbf{Fine-grained understanding on FG-OVD}. All methods using a ViT-B/16 backbone fine-tuned on DOCCI.}
    \label{tab:fg_ovd}
    \resizebox{0.6\columnwidth}{!}{
    \begin{tabular}{l c c c c c}
        \toprule
        \textbf{Method} & \textbf{hard} & \textbf{medium} & \textbf{easy} & \textbf{trivial} & \textbf{Avg} \\
        \midrule
        FineLIP          & 18.17 & 38.68 & 41.96 & \cellcolor{blue!15}73.79 & 43.15 \\
        GOAL             & 18.65 & 39.66 & 44.50 & 72.78                    & \underline{43.90} \\
        \textbf{MulCLIP} & \cellcolor{blue!15}19.24 & \cellcolor{blue!15}40.73 & \cellcolor{blue!15}47.57 & \underline{68.89} & \cellcolor{blue!15}44.11 \\
        W/o SAP          & 16.56 & 37.84 & 43.03 & 65.84                    & 40.82 \\
        W/o WPR          & 17.38 & 38.51 & 45.42 & 68.41                    & 42.43 \\
        \bottomrule
    \end{tabular}}
\end{table}

\section{Fine-grained quantitative result}
\label{app:fg_quantitative_result}
\subsection{Fine-grained Understanding across Difficulty Levels}
While previous experiments primarily assess image-level retrieval, they mainly capture how well a model aligns global scene semantics with long or short descriptions. To explicitly probe local grounding, we further evaluate MulCLIP and baselines on the fine-grained FG-OVD \citep{citeFG-OVD}, which evaluates whether a model can distinguish a correct region-level caption from perturbed captions that modify fine-grained attributes such as color, material, count, and spatial relations.

In FG-OVD, each region is annotated with one positive caption and a set of perturbed negatives created by replacing specific attribute words such as color, material, or spatial relations. These candidates are grouped into four difficulty levels---hard, medium, easy, and trivial---depending on how similar the negatives remain to the positive description, with the hardest cases differing by only one or two attributes. Following the standard protocol, we rank each region's true caption among its candidates. As shown in Tab.~\ref{tab:fg_ovd}, MulCLIP consistently outperforms the other adaptation methods on the hard, medium, and easy splits, confirming that its multi-level alignment enhances sensitivity to subtle attribute changes.

\subsection{Additional Grounding Evaluation on RefCOCO}
\label{app:refcoco_grounding}

To complement the qualitative attention-map analysis in the main paper, we further evaluate grounding performance on standard referring-expression comprehension benchmarks. Specifically, we report zero-shot REC results on RefCOCOg, RefCOCO+, and RefCOCO using the Zeroshot\_REC evaluation protocol. Referring expression comprehension requires a model to localize the target object described by a natural-language expression, and the RefCOCO family is a standard benchmark suite for this task~\citep{refZeroshotREC}.

As shown in Table~\ref{tab:refcoco_grounding}, MulCLIP consistently outperforms GOAL on all eight evaluated splits. The gains are especially clear on RefCOCO+ Test A and RefCOCO Val, where MulCLIP improves over GOAL by 6.36 and 3.32 percentage points, respectively. Since both methods are evaluated under the same Zeroshot\_REC setting, these results suggest that MulCLIP improves spatial grounding beyond visualization effects alone. This provides additional quantitative evidence that multi-level alignment benefits not only long-text retrieval, but also downstream localization fidelity.

\begin{table}[htbp]
\centering
\scriptsize
\setlength{\tabcolsep}{4pt}
\renewcommand{\arraystretch}{1.08}
\caption{\textbf{Zero-shot referring expression comprehension on RefCOCO benchmarks.}
We evaluate MulCLIP and GOAL under the same Zeroshot\_REC baseline protocol. MulCLIP outperforms GOAL on all evaluated splits, suggesting improved spatial grounding and localization fidelity.}
\label{tab:refcoco_grounding}
\resizebox{\linewidth}{!}{%
\begin{tabular}{lcccccccc}
\toprule
\textbf{Method}
& \textbf{RefCOCOg Val}
& \textbf{RefCOCOg Test}
& \textbf{RefCOCO+ Val}
& \textbf{RefCOCO+ Test A}
& \textbf{RefCOCO+ Test B}
& \textbf{RefCOCO Val}
& \textbf{RefCOCO Test A}
& \textbf{RefCOCO Test B} \\
\midrule
GOAL
& 63.26 & 63.35
& 52.88 & 55.22 & 50.38
& 47.44 & 49.37 & 45.95 \\
\textbf{MulCLIP}
& \textbf{65.10} & \textbf{64.16}
& \textbf{56.14} & \textbf{61.58} & \textbf{52.65}
& \textbf{50.76} & \textbf{54.73} & \textbf{46.40} \\
\midrule
\textbf{Gain}
& +1.84 & +0.81
& +3.26 & +6.36 & +2.27
& +3.32 & +5.36 & +0.45 \\
\bottomrule
\end{tabular}%
}

\end{table}

\section{Additional Ablations.}
\label{tab:addition_ablation}
\subsection{Ablations on Word-Patch Reconstruction (WPR) Alignment}
\label{wpa}
\paragraph{Protocol.}
Table~\ref{tab:ablation_short_understanding} compares MulCLIP against three ablated variants that modify the word--
patch objective:

(i) "\textit{MulCLIP w/ Text-recon}" is the full framework but sets \(\mathcal{L}_{\text{Word}}=\mathcal{L}_{\text{recon}}^{\text{text}}\);

(ii) "\textit{MulCLIP w/ Image-recon}" is the full framework but sets \(\mathcal{L}_{\text{Word}}=\mathcal{L}_{\text{recon}}^{\text{image}}\);

(iii) "\textit{MulCLIP w/o Recon}" (We treat the batch-wise token contrastive variant as a naive alternative) is the full framework but replaces token reconstruction with a batch-contrastive alignment between refined tokens and patches, which sets \(\mathcal{L}_{\text{Word}}=\mathcal{L}^{\text{batch}}_{\text{contrast}}(v',\,t')\).
Here \(v'\) and \(t'\) denote refined patch and token embeddings, respectively.

The full bidirectional WPR objective achieves the best average performance for ViT-B/16 and remains competitive for ViT-L/14. Although one-sided variants occasionally perform better on individual metrics, the bidirectional objective provides a more stable trade-off across long- and short-caption retrieval. When reconstruction is restricted to a single direction, the model remains effective on short captions, where one-to-one cues dominate. However, such one-sided objectives and naive approach reveal consistent shortcomings in cross-domain long-text transfer. By contrast, the full bidirectional scheme balances both perspectives and avoids collapsing into a single retrieval path, leading to more stable results under distribution shifts.

\begin{table*}[htbp]
  \centering
  \resizebox{\linewidth}{!}{%
    \begin{tabular}{@{} l l l
                     c c c c  %
                     c c c c  %
                     c c c c  %
                     c c c c  %
                     c        %
                     @{}}
      \toprule
      &  & \textbf{Method}
      & \multicolumn{4}{c}{\textbf{Urban-1K}}
      & \multicolumn{4}{c}{\textbf{DCI}}
      & \multicolumn{4}{c}{\textbf{DOCCI}}
      & \multicolumn{4}{c}{\textbf{COCO}}
      & \textbf{Avg} \\
      \cmidrule(lr){4-7}\cmidrule(lr){8-11}\cmidrule(lr){12-15}\cmidrule(lr){16-19}
      &  &
      & \multicolumn{2}{c}{\textbf{T$\Rightarrow$I}} & \multicolumn{2}{c}{\textbf{I$\Rightarrow$T}}
      & \multicolumn{2}{c}{\textbf{T$\Rightarrow$I}} & \multicolumn{2}{c}{\textbf{I$\Rightarrow$T}}
      & \multicolumn{2}{c}{\textbf{T$\Rightarrow$I}} & \multicolumn{2}{c}{\textbf{I$\Rightarrow$T}}
      & \multicolumn{2}{c}{\textbf{T$\Rightarrow$I}} & \multicolumn{2}{c}{\textbf{I$\Rightarrow$T}}
      & \\
      \cmidrule(lr){4-5}\cmidrule(lr){6-7}
      \cmidrule(lr){8-9}\cmidrule(lr){10-11}
      \cmidrule(lr){12-13}\cmidrule(lr){14-15}
      \cmidrule(lr){16-17}\cmidrule(lr){18-19}
      &  &
      & R@1 & R@5 & R@1 & R@5
      & R@1 & R@5 & R@1 & R@5
      & R@1 & R@5 & R@1 & R@5
      & R@1 & R@5 & R@1 & R@5
      & \textbf{Avg} \\
      \midrule
      & \multirow{4}{*}{\rotatebox{90}{\scriptsize ViT-B/16}}
      & MulCLIP w/ Text-recon
            & 73.80 & 92.00 & 81.10 & 95.40
            & 65.38 & 83.54 & 65.88 & 84.59
            & 82.27 & \underline{97.25} & \underline{80.94} & \cellcolor{blue!15}{96.92}
            & \underline{37.52} & 63.23 & \cellcolor{blue!15}{55.06} & \underline{78.28}
            & 77.07 \\
      &  & MulCLIP w/ Image-recon
            & 73.20 & 91.60 & \underline{80.80} & 94.90
            & \underline{67.08} & \underline{84.89} & 65.73 & \underline{84.79}
            & \underline{82.71} & \cellcolor{blue!15}{97.45} & 80.78 & \cellcolor{blue!15}{96.92}
            & 37.31 & \cellcolor{blue!15}{63.28} & 54.52 & 78.16
            & 77.13 \\
      &  & MulCLIP w/o Reconstruction
            & \underline{73.90} & \underline{92.30} & 80.60 & \underline{95.60}
            & 66.88 & 83.84 & \underline{66.68} & \cellcolor{blue!15}{85.04}
            & \cellcolor{blue!15}{82.73} & 97.00 & \cellcolor{blue!15}{80.96} & 96.90
            & 37.19 & 62.78 & 54.26 & 78.14
            & \underline{77.18} \\
      & & \textbf{MulCLIP (ours)}
            & \cellcolor{blue!15}{77.30} & \cellcolor{blue!15}{92.60} & \cellcolor{blue!15}{84.00} & \cellcolor{blue!15}{96.10}
            & \cellcolor{blue!15}{69.10} & \cellcolor{blue!15}{86.00} & \cellcolor{blue!15}{67.10} & 84.20
            & 82.20 & 97.10 & 80.30 & 96.90
            & \cellcolor{blue!15}{37.58} & \underline{63.26} & \underline{54.76} & \cellcolor{blue!15}{78.64}
            & \cellcolor{blue!15}77.95 \\
      \midrule
      & \multirow{4}{*}{\rotatebox{90}{\scriptsize ViT-L/14}}
      & MulCLIP w/ Text-recon
            & \cellcolor{blue!15}{86.20} & 96.50 & \underline{87.60} & 97.00
            & \cellcolor{blue!15}{73.74} & 87.74 & 71.24 & \underline{87.84}
            & \underline{86.47} & 98.22 & 84.33 & 97.78
            & \underline{43.90} & \underline{69.61} & \underline{61.30} & \underline{83.66}
            & \underline{82.07} \\
      &  & MulCLIP w/ Image-recon
            & 84.80 & \underline{96.90} & 86.60 & \underline{97.20}
            & 73.54 & \cellcolor{blue!15}{88.84} & \cellcolor{blue!15}{72.14} & \cellcolor{blue!15}{88.54}
            & 86.02 & \cellcolor{blue!15}{98.41} & \underline{84.67} & \cellcolor{blue!15}{97.96}
            & \cellcolor{blue!15}{44.02} & 69.54 & \cellcolor{blue!15}{61.50} & \cellcolor{blue!15}{84.02}
            & \cellcolor{blue!15}82.17 \\
      &  & MulCLIP w/o Reconstruction
            & 83.00 & 95.20 & 87.20 & 96.50
            & 72.54 & 87.79 & \underline{71.29} & \underline{87.84}
            & 86.27 & \underline{98.35} & 84.12 & \underline{97.94}
            & 43.48 & 68.86 & 60.60 & 83.24
            & 81.51 \\
      & & \textbf{MulCLIP (ours)}
            & \underline{85.80} & \cellcolor{blue!15}{97.10} & \cellcolor{blue!15}{88.30} & \cellcolor{blue!15}97.30
            & \underline{73.70} & \underline{88.20} & 70.80 & 86.90
            & \cellcolor{blue!15}{86.70} & 98.10 & \cellcolor{blue!15}{84.80} & 97.90
            & 43.69 & \cellcolor{blue!15}{69.73} & 60.76 & 83.00
            & 82.05 \\
      \bottomrule
    \end{tabular}%
  }
    \caption{
Ablation of word--patch alignment variants. All models are fine-tuned on DOCCI under the same protocol. The only modified component is the word--patch objective.
}
 \label{tab:ablation_short_understanding}
\end{table*}

\subsection{Ablation on local-loss weights}

\paragraph{Setup.}
To examine sensitivity to the local losses, we tie the two local weights and sweep
\(\lambda_{\text{word}} = \lambda_{\text{sub}} \in \{0.2, 0.6, 0.8, 1.0\}\) on the ViT-B/16 checkpoint fine-tuned on DOCCI. For each setting, we evaluate R@1 on long-text benchmarks (DOCCI, DCI, Urban1K) and short-text benchmarks (Flickr30K, COCO), as summarized in Table~\ref{tab:lambda_tied}.

\begin{table}[!ht]
  \centering
  \caption{\textbf{Ablation of tied local-loss weight \(\lambda_{\text{word}} = \lambda_{\text{sub}}\) (ViT-B/16, DOCCI FT).}
  We report R@1 (\%) for text-to-image (T$\Rightarrow$I) and image-to-text (I$\Rightarrow$T) retrieval on long-text (DOCCI, DCI, Urban1K) and short-text (Flickr30K, COCO) benchmarks.}
  \label{tab:lambda_tied}
  \scriptsize
  \setlength{\tabcolsep}{4pt}
  \renewcommand{\arraystretch}{1.05}
  \begin{tabular}{@{} c
                  c c c c c
                  c c c c c
                  c @{}}
    \toprule
    \multirow{2}{*}{$\lambda_{\text{word}} = \lambda_{\text{sub}}$}
      & \multicolumn{5}{c}{\textbf{Text-to-Image R@1 (\%)}}
      & \multicolumn{5}{c}{\textbf{Image-to-Text R@1 (\%)}}
      & \textbf{Avg} \\
    \cmidrule(lr){2-6}\cmidrule(lr){7-11}
      & DOCCI & DCI & Urban1K & Flickr30K & COCO
      & DOCCI & DCI & Urban1K & Flickr30K & COCO
      &  \\
    \midrule
    0.2 & 82.2 & 66.9 & 72.6 & 67.1 & 37.4
        & 80.3 & 64.1 & 82.0 & \textbf{84.4} & \textbf{55.1}
        & 69.21 \\
    0.6 & \textbf{82.6} & 67.3 & 74.0 & 66.8 & 37.6
        & \textbf{80.7} & 66.1 & 81.9 & 82.2 & 54.8
        & 69.40 \\
    0.8 & 82.2 & 66.3 & 72.2 & 66.7 & 37.5
        & \textbf{80.7} & 65.2 & 82.0 & 81.4 & 54.8
        & 68.90 \\
    1.0 & 82.2 & \textbf{69.1} & \textbf{77.3} & \textbf{67.4} & \textbf{37.7}
        & 80.3 & \textbf{67.1} & \textbf{84.0} & 81.9 & 54.8
        & \textbf{70.18} \\
    \bottomrule
  \end{tabular}
\end{table}

When we vary \(\lambda_{\text{word}} = \lambda_{\text{sub}}\) from 0.2 to 1.0, long-text R@1 varies by up to 4.3 percentage points and short-text R@1 by up to 3.0 points. In-domain performance on DOCCI is almost flat, while DCI and Urban1K show mild gains as \(\lambda\) increases. Our default choice \(\lambda=1.0\) slightly favors long-text retrieval (especially on DCI and Urban1K) without noticeably degrading short-text performance. Overall, these results indicate that MulCLIP is robust with respect to the local-loss weights within a broad mid-range.

\subsection{Robustness to number of subcaptions}

\paragraph{Setup.}
We study how sensitive MulCLIP is to the number of sentence-level subcaptions. We fine-tune ViT-B/16 on DOCCI while varying the maximum number of sentences per caption from 5 to 20, and evaluate R@1 on long-text (DOCCI, DCI, Urban1K) and short-text (Flickr30K, COCO) retrieval. Subcaptions are defined at the sentence level using punctuation-based splitting.

\begin{table}[ht]
  \centering
  \caption{\textbf{Effect of caption granularity (max sentences per caption).}
  R@1 (\%) for text-to-image (T$\Rightarrow$I) and image-to-text (I$\Rightarrow$T) retrieval on long-text (DOCCI, DCI, Urban1K) and short-text (Flickr30K, COCO) benchmarks.}
  \label{tab:max_sentences}
  \scriptsize
  \setlength{\tabcolsep}{4pt}
  \renewcommand{\arraystretch}{1.05}
  \begin{tabular}{@{} c
                  c c c c c
                  c c c c c
                  c @{}}
    \toprule
    \multirow{2}{*}{Max sentences}
      & \multicolumn{5}{c}{\textbf{Text-to-Image R@1 (\%)}}
      & \multicolumn{5}{c}{\textbf{Image-to-Text R@1 (\%)}}
      & \textbf{Avg} \\
    \cmidrule(lr){2-6}\cmidrule(lr){7-11}
      & DOCCI & DCI & Urban1K & Flickr30K & COCO
      & DOCCI & DCI & Urban1K & Flickr30K & COCO
      &  \\
    \midrule
     5  & \textbf{82.9} & 66.4 & 76.2 & 65.6 & 37.1
        & 81.1 & 65.5 & 81.2 & 81.5 & 54.2
        & 69.17 \\
    10  & 82.3 & 65.8 & 74.8 & 66.3 & 37.3
        & \textbf{81.4} & 66.7 & 82.1 & 81.9 & \textbf{54.5}
        & 69.31 \\
    15 (default)
        & 82.6 & 67.5 & 75.3 & 66.1 & 36.9
        & 80.9 & 66.5 & 81.1 & \textbf{82.3} & 54.3
        & 69.35 \\
    20  & 82.2 & \textbf{69.0} & \textbf{77.5} & \textbf{67.4} & \textbf{37.9}
        & 80.3 & \textbf{67.1} & \textbf{83.9} & 81.9 & 53.8
        & \textbf{70.10} \\
    \bottomrule
  \end{tabular}
\end{table}

\begin{figure}[ht]
  \centering
  \includegraphics[width=\linewidth]{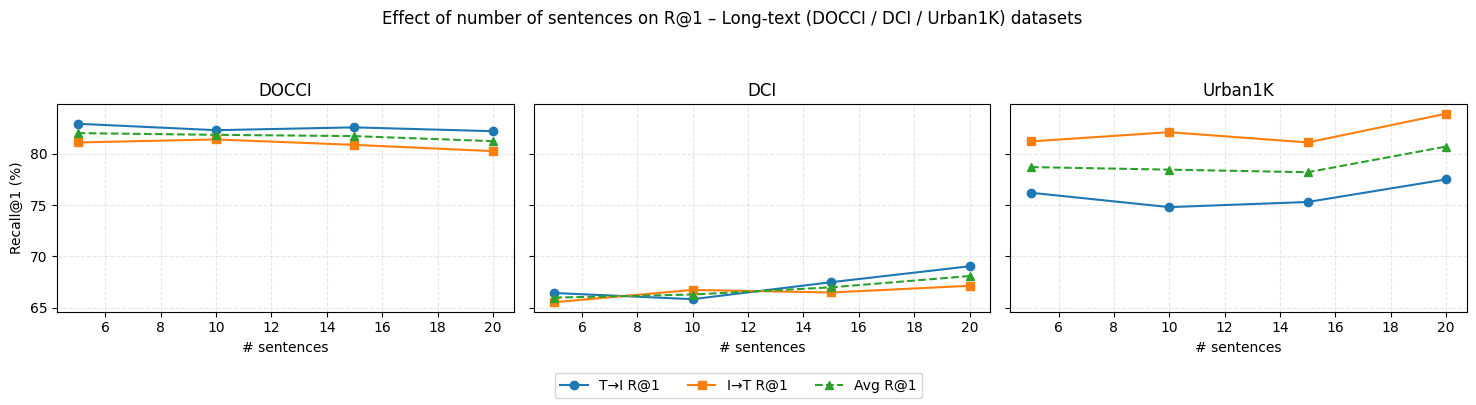}
  \caption{Effect of maximum number of sentences on long-text retrieval (DOCCI / DCI / Urban1K).}
  \label{fig:sent_long}
\end{figure}

\begin{figure}[ht]
  \centering
  \includegraphics[width=\linewidth]{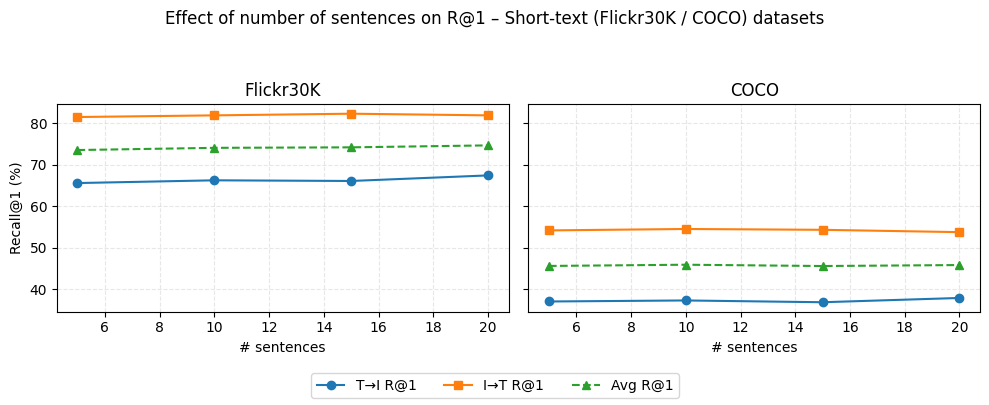}
  \caption{Effect of maximum number of sentences on short-text retrieval (Flickr30K / COCO).}
  \label{fig:sent_short}
\end{figure}

\begin{figure}[ht]
  \centering
  \includegraphics[width=0.75\linewidth]{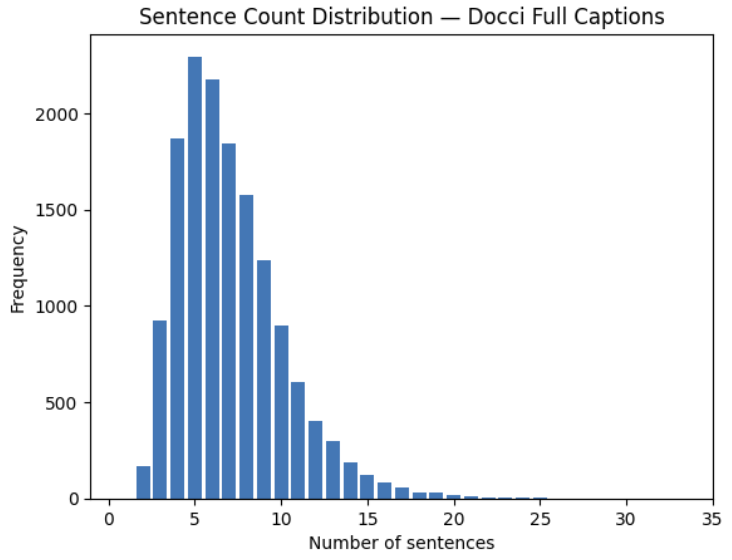}
  \caption{Sentence count distribution on DOCCI.}
  \label{fig:docci_sent_dist}
\end{figure}

\begin{figure}[ht]
  \centering
  \includegraphics[width=0.75\linewidth]{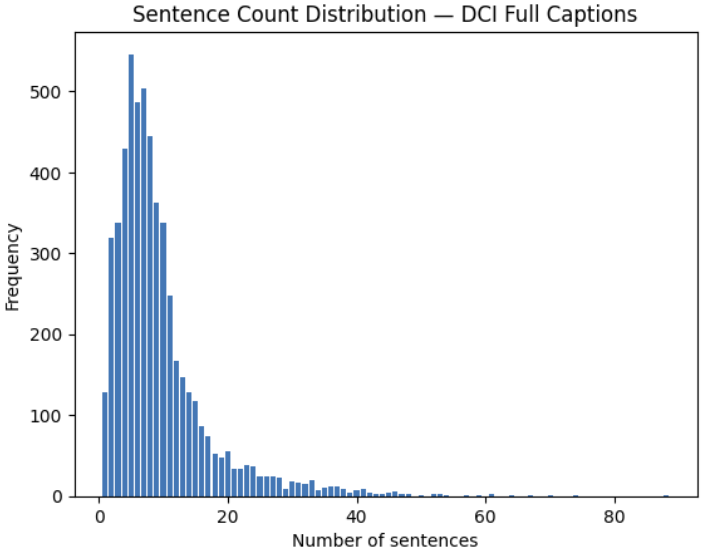}
  \caption{Sentence count distribution on DCI.}
  \label{fig:dci_sent_dist}
\end{figure}

Across the range from 5 to 20 sentences, short-text R@1 on Flickr30K and COCO remains almost flat. DOCCI and DCI show small gains when increasing from very few sentences to around 10--20, after which performance saturates. Urban1K shows a mild upward trend as more subcaptions are retained, and no setting exhibits a severe degradation.

Sentence-count histograms for DOCCI and DCI (Figures~\ref{fig:docci_sent_dist}--\ref{fig:dci_sent_dist}) show that most captions contain 3--10 sentences, with only a small fraction exceeding 20. Thus, our default cap of 15 sentences typically includes all available sentences without over-fragmenting the caption. Overall, MulCLIP benefits from multiple sentence-level subcaptions but remains stable across a wide range of reasonable caps, indicating robustness to caption granularity.

\subsection{Sensitivity to sentence-level subcaption splitting.}

\begin{figure}[t]
    \centering
    \includegraphics[width=0.7\linewidth]{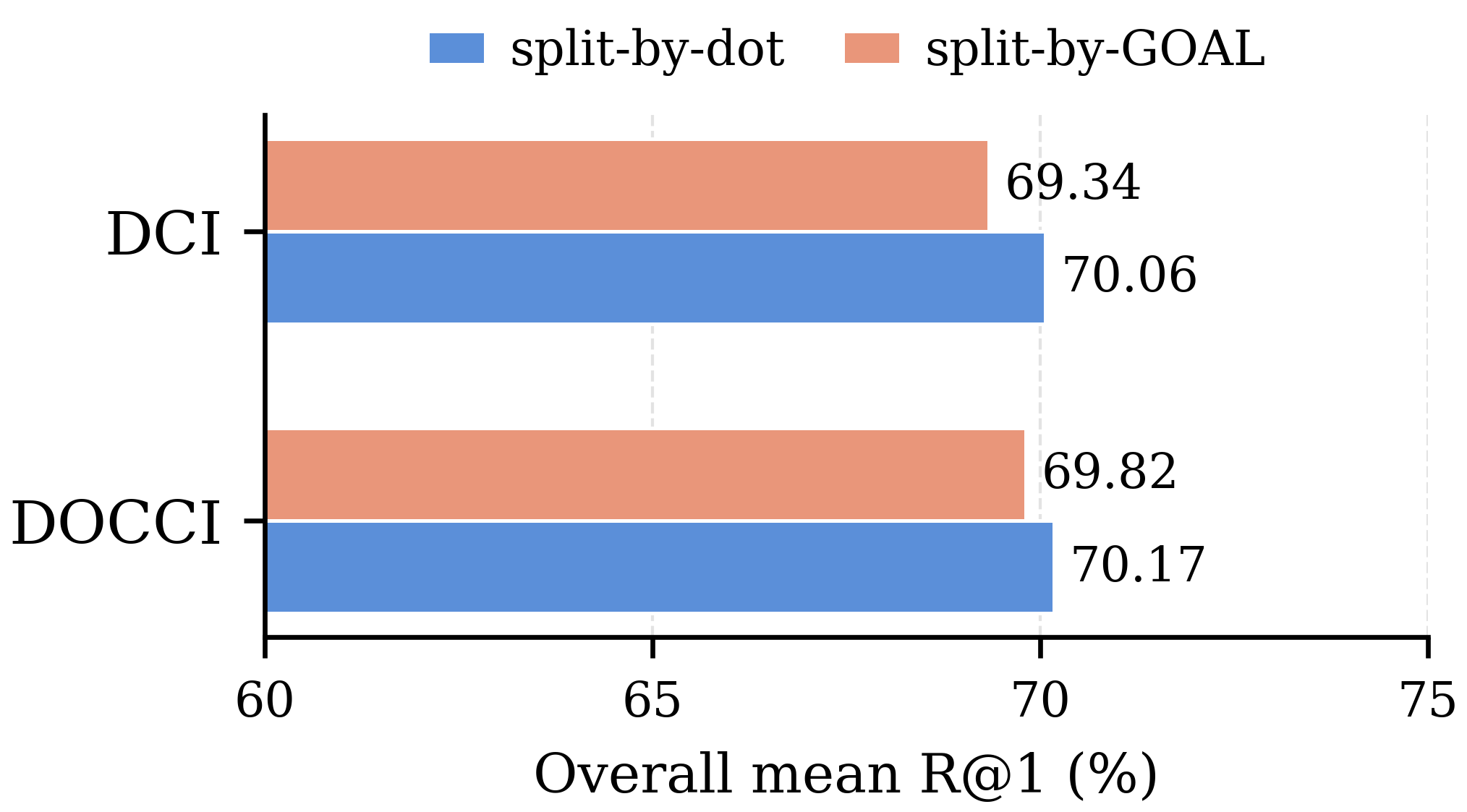}
    \caption{\textbf{Ablation on sentence-level subcaption splitting.}
    We compare the default punctuation-based splitting strategy (\textbf{split-by-dot}) with a GOAL-style filtered-caption construction (\textbf{split-by-GOAL}). Results are averaged over the evaluated retrieval benchmarks for each fine-tuning setting. The close performance suggests that SAP remains stable under different subcaption construction strategies.}
    \label{fig:subcaption_splitting}
\end{figure}

We further examine whether SAP is sensitive to the way sentence-level subcaptions are constructed. In addition to our default punctuation-based splitting (\textbf{split-by-dot}), we evaluate a GOAL-style filtered-caption variant (\textbf{split-by-GOAL}) while keeping the backbone, training data, and optimization protocol unchanged. As shown in Fig.~\ref{fig:subcaption_splitting}, the two variants produce close retrieval performance across both DOCCI and DCI fine-tuning settings. For DOCCI fine-tuning, split-by-dot obtains an overall mean R@1 of 70.18, compared with 69.82 for split-by-GOAL. For DCI fine-tuning, the corresponding scores are 70.06 and 69.34. These results suggest that, on structured descriptive benchmarks, SAP is not overly sensitive to the exact sentence-level subcaption construction.

\subsection{Fair comparison with FineLIP and role of the global loss}

\paragraph{Total loss and local modules.}
For clarity, the full MulCLIP objective can be written as
\begin{equation}
\mathcal{L}_{\text{total}}
= \mathcal{L}_{\text{global}}\bigl(v_{\text{cls}}, t^{\text{long}}_{\text{eot}}, t^{\text{short}}_{\text{eot}}\bigr)
+ \lambda_{\mathrm{W}} \,\mathcal{L}_{\text{Word}}\bigl(v', t'\bigr)
+ \lambda_{\mathrm{S}} \,\mathcal{L}_{\text{Sub}}\bigl(v', t^{\text{sub}}_{\text{eot}}\bigr),
\end{equation}
where \(v_{\text{cls}}\) is the global image embedding, \(t^{\text{long}}_{\text{eot}}, t^{\text{short}}_{\text{eot}}\) are global text embeddings for long and short captions, and \(v', t'\) are locally calibrated tokens used by the word- and subcaption-level objectives.

\paragraph{Removing the global objective.}
To isolate the contribution of our local modules, we train a variant that \emph{removes} the global loss and keeps only local alignment:
\begin{equation}
\mathcal{L}_{\text{total}}
= \mathcal{L}_{\text{Word}}(v', t')
+ \mathcal{L}_{\text{Sub}}(v', t_{\text{sub}}).
\end{equation}
As shown in Table~\ref{tab:no_global}, this ``No Global'' model suffers a large drop on all three long-text benchmarks compared to full MulCLIP, with R@1 roughly halved in many cases. This confirms that local objectives alone are not sufficient for robust long-text understanding, and that they must work \emph{together} with a strong global alignment term.

\begin{table}[ht]
  \centering
  \caption{\textbf{Effect of removing the global loss (ViT-B/16, DOCCI FT).}
  R@1 (\%) for text-to-image (T$\Rightarrow$I) and image-to-text (I$\Rightarrow$T) retrieval.}
  \label{tab:no_global}
  \scriptsize
  \setlength{\tabcolsep}{5pt}
  \renewcommand{\arraystretch}{1.05}
  \begin{tabular}{@{} l
                  c c c c c c c @{}}
    \toprule
    \textbf{Method}
      & \multicolumn{2}{c}{\textbf{DCI}}
      & \multicolumn{2}{c}{\textbf{DOCCI}}
      & \multicolumn{2}{c}{\textbf{Urban1K}}
      & \textbf{Avg} \\
    \cmidrule(lr){2-3}\cmidrule(lr){4-5}\cmidrule(lr){6-7}
      & T$\Rightarrow$I & I$\Rightarrow$T
      & T$\Rightarrow$I & I$\Rightarrow$T
      & T$\Rightarrow$I & I$\Rightarrow$T
      & \\
    \midrule
    MulCLIP (full)
      & \textbf{69.1} & \textbf{67.1}
      & \textbf{82.2} & \textbf{80.3}
      & \textbf{77.3} & \textbf{84.0}
      & \textbf{76.67} \\
    No Global
      & 33.47 & 28.86
      & 40.25 & 29.41
      & 25.00 & 36.30
      & 32.22 \\
    \bottomrule
  \end{tabular}
\end{table}

\paragraph{Adding our global loss to FineLIP.}
We next equip FineLIP with the \emph{same} long/short global objective and 50\% token compression as MulCLIP. Let
\begin{equation}
V = v' \oplus v_{\text{cls}}, \quad
T = t' \oplus t_{\text{eot}}
\end{equation}
be the concatenation of global and local tokens. The original FineLIP paper provides two runnable variants of its triplet-based CLIM/FILIP objective \(R(\cdot)\): \(R(V, T)\) and \(R(v', t')\). We therefore define:

\noindent\textbf{FineLIP ver.~1 (+Global):}
\begin{equation}
\mathcal{L}_{\text{total}}
= \mathcal{L}_{\text{global}}(v_{\text{cls}}, t^{\text{long}}_{\text{eot}}, t^{\text{short}}_{\text{eot}})
+ R(V, T),
\end{equation}

\noindent\textbf{FineLIP ver.~2 (+Global):}
\begin{equation}
\mathcal{L}_{\text{total}}
= \mathcal{L}_{\text{global}}(v_{\text{cls}}, t^{\text{long}}_{\text{eot}}, t^{\text{short}}_{\text{eot}})
+ R(v', t').
\end{equation}

For a fair comparison, we match these two FineLIP variants against our \emph{W/o SAP} model
\begin{equation}
\mathcal{L}_{\text{total}}
= \mathcal{L}_{\text{global}}(v_{\text{cls}}, t^{\text{long}}_{\text{eot}}, t^{\text{short}}_{\text{eot}})
+ \mathcal{L}_{\text{Word}}(v', t'),
\end{equation}
so all methods share the same backbone, global loss, and token-compression ratio, differing only in how local interactions are modeled (FineLIP's CLIM/FILIP vs.\ our Word--Patch Reconstruction).

All three models are fine-tuned on DOCCI under the same protocol and evaluated on Urban1K, DCI, and DOCCI. Tables~\ref{tab:finelip_fair_style} summarize R@1 for both directions.

\begin{table}[ht!]
  \centering
  \caption{\textbf{Fair comparison between FineLIP+Global and our Global+LC+WPR (W/o SAP)}.
  All models share the same backbone, global loss, token-compression ratio, and DOCCI fine-tuning protocol.
  Best score per backbone is highlighted.}
  \label{tab:finelip_fair_style}
  \resizebox{0.90\linewidth}{!}{%
  \begin{tabular}{@{} l l l
      c c
      c c
      c c
      c @{}}
    \toprule
     & & \textbf{Method}
     & \multicolumn{2}{c}{\textbf{Urban1K}}
     & \multicolumn{2}{c}{\textbf{DCI}}
     & \multicolumn{2}{c}{\textbf{DOCCI}}
     & \textbf{Avg} \\
    \cmidrule(lr){4-5}\cmidrule(lr){6-7}\cmidrule(lr){8-9}
     & &
     & T$\Rightarrow$I & I$\Rightarrow$T
     & T$\Rightarrow$I & I$\Rightarrow$T
     & T$\Rightarrow$I & I$\Rightarrow$T
     &  \\
    \midrule

\multirow{3}{*}{\rotatebox{90}{\scriptsize ViT-B/16}}
  & & FineLIP (ver.~1, +Global)
      & 64.9 & 71.4
      & 56.4 & 44.8
      & 65.6 & 60.0
      & 60.5 \\

  & & FineLIP (ver.~2, +Global)
      & 64.9 & 74.4
      & 56.2 & 44.7
      & 65.5 & 59.7
      & 60.9 \\

  & & \textbf{Ours --- W/o SAP}
      & \textbf{74.4} & \textbf{80.1}
      & \textbf{65.4} & \textbf{64.1}
      & \textbf{80.6} & \textbf{78.9}
      & \textbf{73.9} \\

\cmidrule(lr){1-10}

\multirow{3}{*}{\rotatebox{90}{\scriptsize ViT-L/14}}
  & & FineLIP (ver.~1, +Global)
      & 68.6 & 73.2
      & 59.6 & 43.3
      & 72.9 & 67.0
      & 64.1 \\

  & & FineLIP (ver.~2, +Global)
      & 65.5 & 73.9
      & 57.7 & 48.0
      & 72.4 & 67.0
      & 64.1 \\

  & & \textbf{Ours --- W/o SAP}
      & \textbf{85.0} & \textbf{87.3}
      & \textbf{71.9} & \textbf{68.5}
      & \textbf{85.8} & \textbf{83.6}
      & \textbf{80.4} \\

    \bottomrule
  \end{tabular}}
\end{table}

Under a fully matched setup (same backbone, global loss, token compression, data, and optimization), both FineLIP+Global variants remain consistently below our Global+LC+WPR (W/o SAP) model on all three long-text benchmarks, in both directions and for both ViT-B/16 and ViT-L/14. Since the only difference is how local tokens are used, this indicates that our Local Calibration and Word--Patch Reconstruction modules exploit compressed local tokens more effectively than FineLIP's CLIM/FILIP interaction.

The comparison with the ``No Global'' variant further highlights the complementarity of components: the global objective is essential for long-text robustness, while LC+WPR provide the additional fine-grained gains on top. In the main ablations, adding SAP on top of Global+LC+WPR then yields further, stable improvements, suggesting that subcaption--patch alignment is complementary rather than the sole driver of MulCLIP's benefits.
\section{Extended Retrieval Qualitative Results}

\begin{figure*}[t]
  \centering

  \begin{minipage}{0.48\textwidth}
    \centering
    \vspace{0pt}

    \begin{minipage}{0.45\textwidth}
      \centering
      \includegraphics[width=\textwidth]{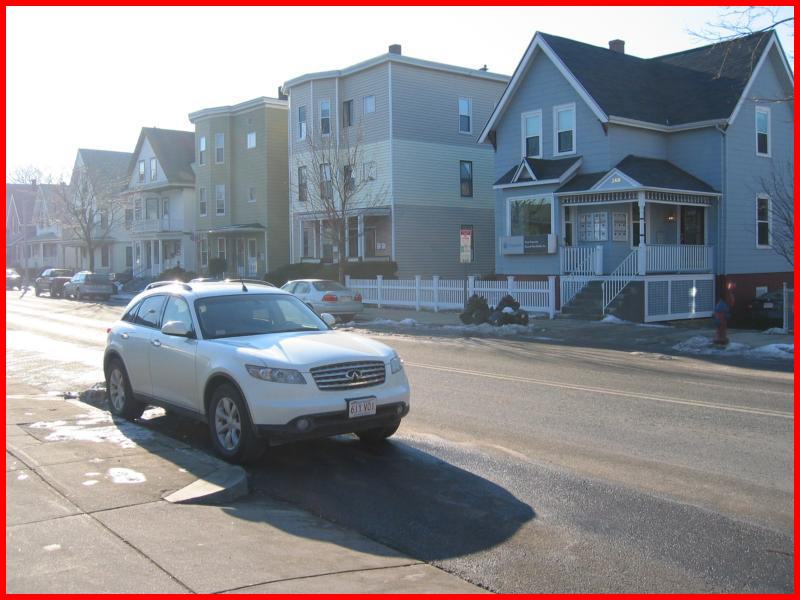}
      \vspace{2pt}
      {\scriptsize \textbf{Retrieved by GOAL}}
    \end{minipage}
    \hfill
    \begin{minipage}{0.45\textwidth}
      \centering
      \includegraphics[width=\textwidth]{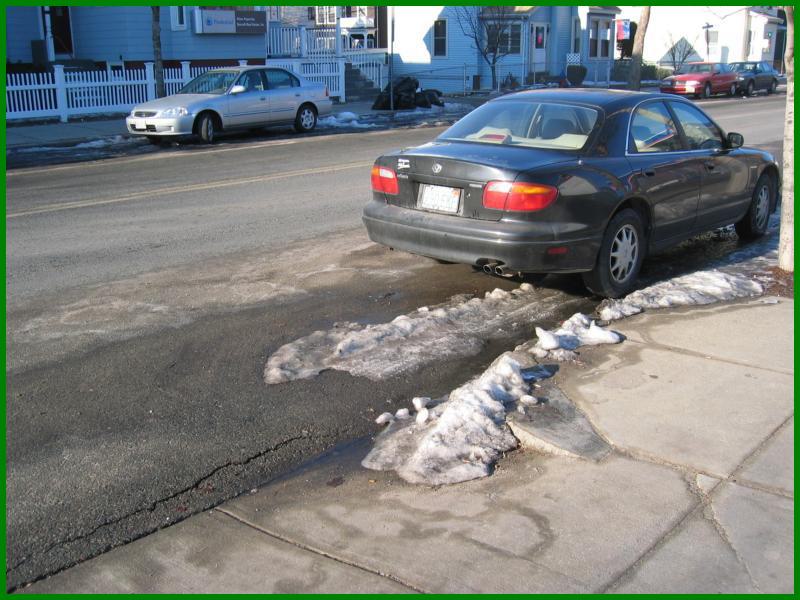}
      \vspace{2pt}
      {\scriptsize \textbf{Retrieved by MulCLIP}}
    \end{minipage}

    \vspace{8pt}
    \begin{minipage}{\textwidth}
      \scriptsize
      \textbf{Query:} \textit{A \colorbox{yellow!30}{dark-colored sedan} is parked askew on the side of a street, half on the asphalt road and half on the concrete sidewalk. Patches of melting snow are present, indicating recent snowfall or wintry conditions. In the background, a white sedan is parked correctly on the opposite side of the street. Residential buildings with white fences, bare deciduous trees, and other parked cars line the street. There's a general sense of a suburban or residential neighborhood on a clear day with sunlight casting shadows on the ground. \colorbox{yellow!30}{Visible tire tracks through the snow} suggest recent vehicle movement.}
    \end{minipage}

  \end{minipage}
  \hfill
  \begin{minipage}{0.48\textwidth}
    \centering
    \vspace{0pt}

    \begin{minipage}{0.45\textwidth}
      \centering
      \includegraphics[width=\textwidth]{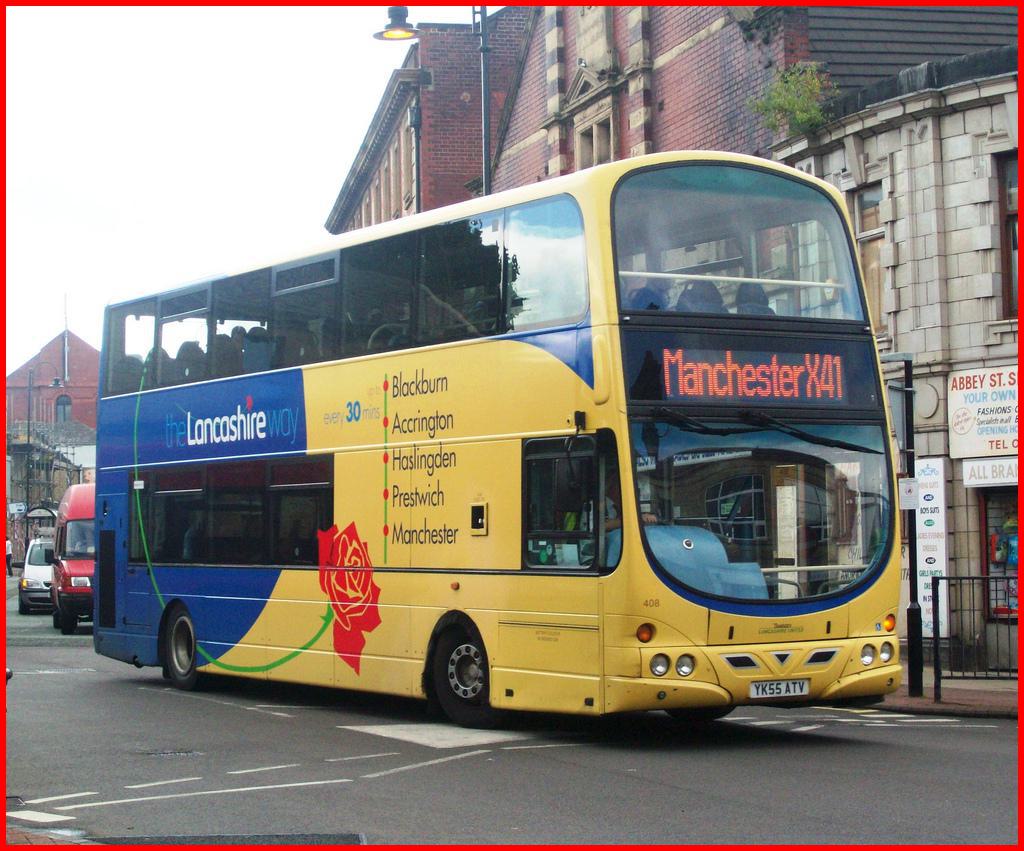}
      \vspace{2pt}
      {\scriptsize \textbf{Retrieved by GOAL}}
    \end{minipage}
    \hfill
    \begin{minipage}{0.45\textwidth}
      \centering
      \includegraphics[width=\textwidth]{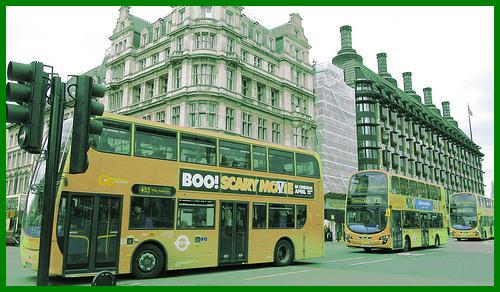}
      \vspace{2pt}
      {\scriptsize \textbf{Retrieved by MulCLIP}}
    \end{minipage}

    \vspace{8pt}
    \begin{minipage}{\textwidth}
      \scriptsize
      \textbf{Query:} \textit{The image displays a vibrant urban scene with \colorbox{yellow!30}{two} modern double-decker buses on a road, presumably in a city in the United Kingdom. The bus in the foreground is painted in a bright yellow color with bold advertising on its side, while \colorbox{yellow!30}{the bus in the background is also yellow} with visible route information. Traffic lights appear on the left, indicating a crosswalk or intersection. European-style architecture is prominent, with elaborate stone buildings adorned with numerous windows and ornamental details. The sky is overcast with hints of \colorbox{yellow!30}{blue peeking through the clouds}, suggesting a typical cloudy day. The greenery of trees is also visible, adding a touch of nature to the urban environment.}
    \end{minipage}

  \end{minipage}

  \vspace{20pt}

  \begin{minipage}{0.48\textwidth}
    \centering
    \vspace{0pt}

    \begin{minipage}{0.45\textwidth}
      \centering
      \includegraphics[width=\textwidth]{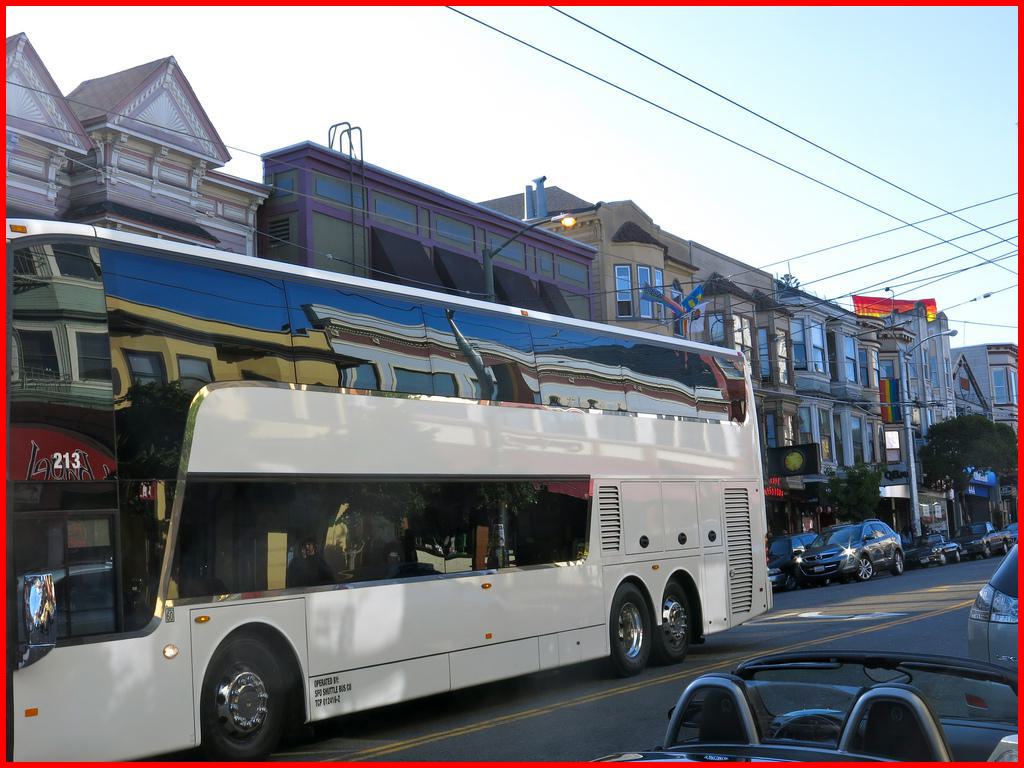}
      \vspace{2pt}
      {\scriptsize \textbf{Retrieved by GOAL}}
    \end{minipage}
    \hfill
    \begin{minipage}{0.45\textwidth}
      \centering
      \includegraphics[width=\textwidth]{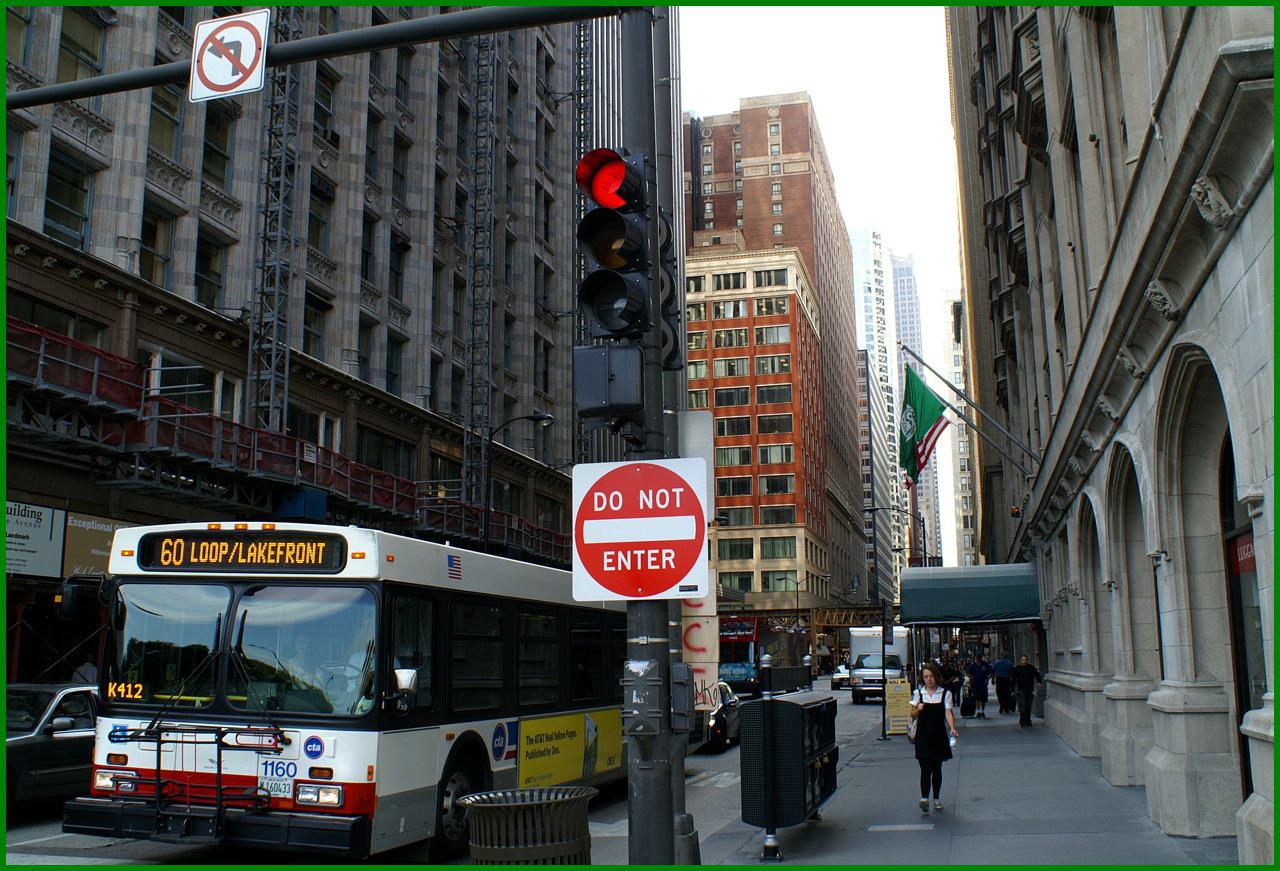}
      \vspace{2pt}
      {\scriptsize \textbf{Retrieved by MulCLIP}}
    \end{minipage}

    \vspace{8pt}
    \begin{minipage}{\textwidth}
      \scriptsize
      \textbf{Query:} \textit{The image depicts a bustling city street scene with clear skies above. In the foreground, a white bus with a digital sign that reads \colorbox{yellow!30}{'60 LOOP/LAKEFRONT'} stops near a \colorbox{yellow!30}{sidewalk, marked 'K412'}. \colorbox{yellow!30}{A red traffic light} hangs above, while a \colorbox{yellow!30}{'DO NOT ENTER' sign} is prominently displayed on a post below. The architecture includes tall, ornamented stone buildings indicative of early 20th-century design, with one building featuring a scaffolding structure along its facade. A \colorbox{yellow!30}{pedestrian} crosses the street, another walks on the sidewalk, and a few flags, including a green, white, and red one, are visible hanging from a building. The urban environment suggests a downtown district, possibly in a large metropolitan city.}
    \end{minipage}

  \end{minipage}
  \hfill
  \begin{minipage}{0.48\textwidth}
    \centering
    \vspace{0pt}

    \begin{minipage}{0.45\textwidth}
      \centering
      \includegraphics[width=\textwidth]{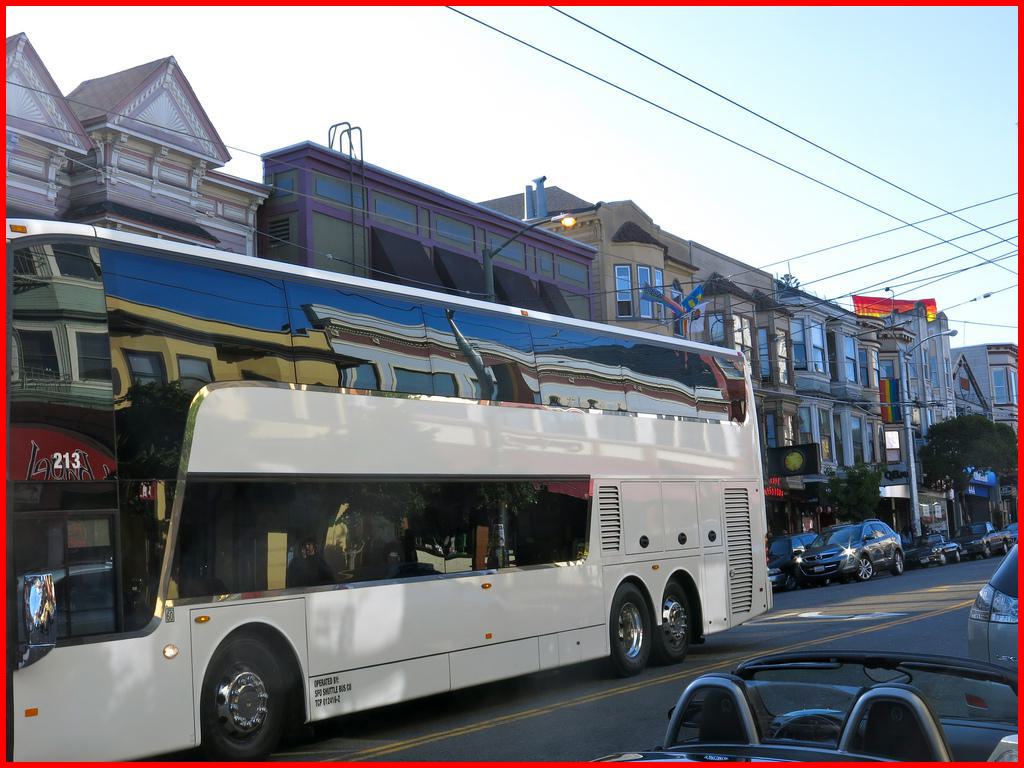}
      \vspace{2pt}
      {\scriptsize \textbf{Retrieved by GOAL}}
    \end{minipage}
    \hfill
    \begin{minipage}{0.45\textwidth}
      \centering
      \includegraphics[width=\textwidth]{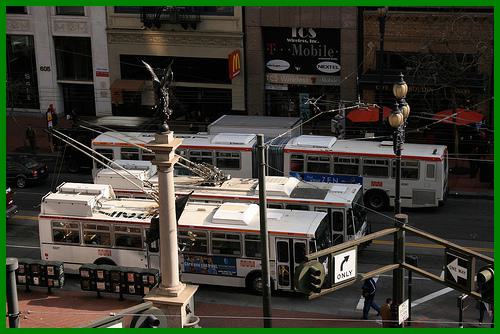}
      \vspace{2pt}
      {\scriptsize \textbf{Retrieved by MulCLIP}}
    \end{minipage}

    \vspace{8pt}
    \begin{minipage}{\textwidth}
      \scriptsize
      \textbf{Query:} \textit{The image captures a busy urban street scene with \colorbox{yellow!30}{two} white articulated trolleybuses, \colorbox{yellow!30}{featuring blue and red stripes}, connected to overhead wires. Above the buses, a \colorbox{yellow!30}{streetlight} with a dual-globe design is visible. In the foreground, \colorbox{yellow!30}{a pole topped with a flying eagle} statue anchors the composition. Behind the buses, several red and white cars are parked. The backdrop is lined with multistory buildings hosting various stores with visible signage. \colorbox{yellow!30}{Pedestrians} can be seen walking along the sidewalks, and traffic lights are located at the street's intersection. The photo, taken from an elevated angle during daylight, shows the street intersecting leftward, with designated lanes for different directions.}
    \end{minipage}

  \end{minipage}

  \vspace{12pt}

  \begin{minipage}{\textwidth}
    \centering
    \scriptsize
    \textbf{Legend:}
    \fcolorbox{green}{white}{\rule{8pt}{8pt}} Correct retrieval \quad
    \fcolorbox{red}{white}{\rule{8pt}{8pt}} Incorrect retrieval \quad
    \colorbox{yellow!30}{Highlighted text} indicates visual details missing from GOAL's retrieved image but correctly matched by MulCLIP
  \end{minipage}

  \caption{Qualitative comparison of text-to-image retrieval between GOAL and MulCLIP. Each pair shows retrieved images from both models for the same query. Colored borders indicate correctness (green: correct; red: incorrect). Yellow highlights denote visual details missing from GOAL's retrieved image but correctly matched by MulCLIP.}
  \label{fig:qualitative_comparison}
\end{figure*}

Fig. ~\ref{fig:qualitative_comparison} and Table~\ref{tab:image_2_text} provide qualitative examples of text-to-image and image-to-text retrieval. These examples illustrate cases in which \textbf{GOAL} retrieves visually plausible but detail-mismatched results, especially when the query depends on small or low-contrast cues such as route numbers, faint text, background signs, small logos, or subtle object attributes. In the displayed cases, \textbf{MulCLIP} retrieves images or captions that better preserve these fine-grained details, which is consistent with the quantitative gains observed on Urban1K and other urban retrieval benchmarks.

\paragraph{Selection protocol.}
We select examples from cases where \textbf{MulCLIP} retrieves the ground-truth item at rank 1 while \textbf{GOAL} does not, and where the query contains visually verifiable fine-grained attributes. These examples are intended to illustrate representative failure modes and should be interpreted as qualitative evidence rather than standalone quantitative proof.

\clearpage

\begin{table}[ht!]
  \centering
  \resizebox{\linewidth}{!}{
  \begin{tabular}{
    m{3.5cm}   %
    m{6cm}     %
    m{6cm}     %
  }
    \toprule
    \textbf{Image query}
    & \textbf{MulCLIP}
    & \textbf{GOAL} \\
    \midrule

    \begin{adjustbox}{valign=t}
      \makebox[\linewidth][c]{%
        \includegraphics[width=3cm,height=2.5cm]{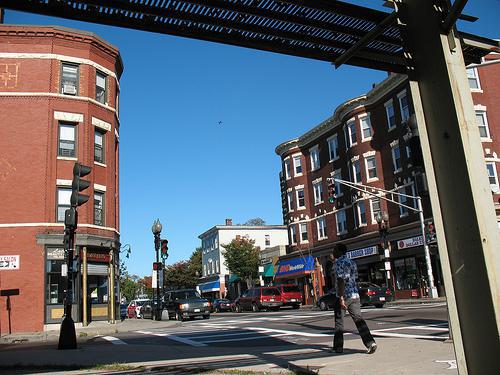}%
      }
    \end{adjustbox}
    &
    \begin{adjustbox}{valign=t}
      \begin{tcolorbox}[
        colback=white,
        colframe=green!70!black,
        boxrule=1pt,
        arc=2pt,
        left=2pt,
        right=2pt,
        top=2pt,
        bottom=2pt,
        width=\linewidth
      ]
      \footnotesize
      This image depicts a vibrant urban street corner on a clear day with blue skies. A person in a blue checkered shirt and casual pants crosses the road at a pedestrian crosswalk, heading towards a series of red brick buildings with rounded and straight corners. The architecture suggests a charming, historic neighborhood with storefronts on the ground level, including one with a blue awning. There's a green streetlight visible and a black lamp post, adding to the quaint ambiance. Vehicles are stopped at the intersection, and the scene includes an overhanging metal structure that could be part of a bus stop. The overall atmosphere is that of a peaceful, sunny day in a bustling city neighborhood.
      \end{tcolorbox}
    \end{adjustbox}
    &
    \begin{adjustbox}{valign=t}
      \begin{tcolorbox}[
        colback=white,
        colframe=red!70!black,
        boxrule=1pt,
        arc=2pt,
        left=2pt,
        right=2pt,
        top=2pt,
        bottom=2pt,
        width=\linewidth
      ]
      \footnotesize
      The image depicts an urban street scene during daytime. In the foreground, \hl{two} individuals with their backs to the camera are walking, one with \hl{a long ponytail and a white shirt, the other with a patterned blouse, and a red backpack}. A \hl{silver car} is visible on the left side of the road, which is marked with multiple round blue traffic signs, indicating no waiting or no stopping restrictions. Across the street, there's a red-bricked five-story building with white stone trimmings and arched windows on the ground floor. The windows on the upper floors are regularly spaced, and the uppermost story appears to be an attic with smaller windows. The sky is clear, suggesting favorable weather conditions.
      \end{tcolorbox}
    \end{adjustbox}
    \\[1em]
    \midrule

    \begin{adjustbox}{valign=t}
      \makebox[\linewidth][c]{%
        \includegraphics[width=3cm,height=2.5cm]{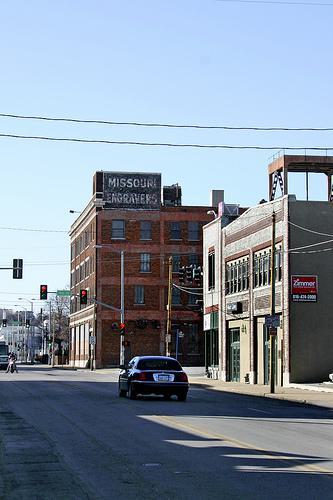}%
      }
    \end{adjustbox}
    &
    \begin{adjustbox}{valign=t}
      \begin{tcolorbox}[
        colback=white,
        colframe=green!70!black,
        boxrule=1pt,
        arc=2pt,
        left=2pt,
        right=2pt,
        top=2pt,
        bottom=2pt,
        width=\linewidth
      ]
      \footnotesize
      The image shows an urban street scene under a clear blue sky. In the foreground, a black car is driving down the road, which is lined with electrical wires above. The architecture is a mix of multi-story brick buildings with visible signs of wear and faded paint, suggesting a historical urban area. The buildings vary in height, with some having flat facades and others featuring recessed windows and ornate detailing. The street appears to be relatively empty, with no pedestrians visible and minimal vehicular traffic. The structures' color palette is predominantly red brick, with accents of tan and white on the secondary building elements.
      \end{tcolorbox}
    \end{adjustbox}
    &
    \begin{adjustbox}{valign=t}
      \begin{tcolorbox}[
        colback=white,
        colframe=red!70!black,
        boxrule=1pt,
        arc=2pt,
        left=2pt,
        right=2pt,
        top=2pt,
        bottom=2pt,
        width=\linewidth
      ]
      \footnotesize
      The image captures a daytime scene on a city street named \hl{"Main Street,"} indicated by a street sign hanging above. Vehicles, including a \hl{red sedan} in the foreground, are \hl{parked along one side of the street, while others, including white vans, are visible in motion}. Pedestrians are present on the sidewalks, some standing and others seated beside buildings; a group congregates near an American flag. Utility poles, traffic signals, and signs, including one indicating a ``Drug-Free School Zone,'' dot the streetscape. Overhead, a concrete overpass spans the thoroughfare. The sky is slightly overcast, casting even lighting across the urban environment.
      \end{tcolorbox}
    \end{adjustbox}
    \\[1em]
    \midrule

    \begin{adjustbox}{valign=t}
      \makebox[\linewidth][c]{%
        \includegraphics[width=3cm,height=2.5cm]{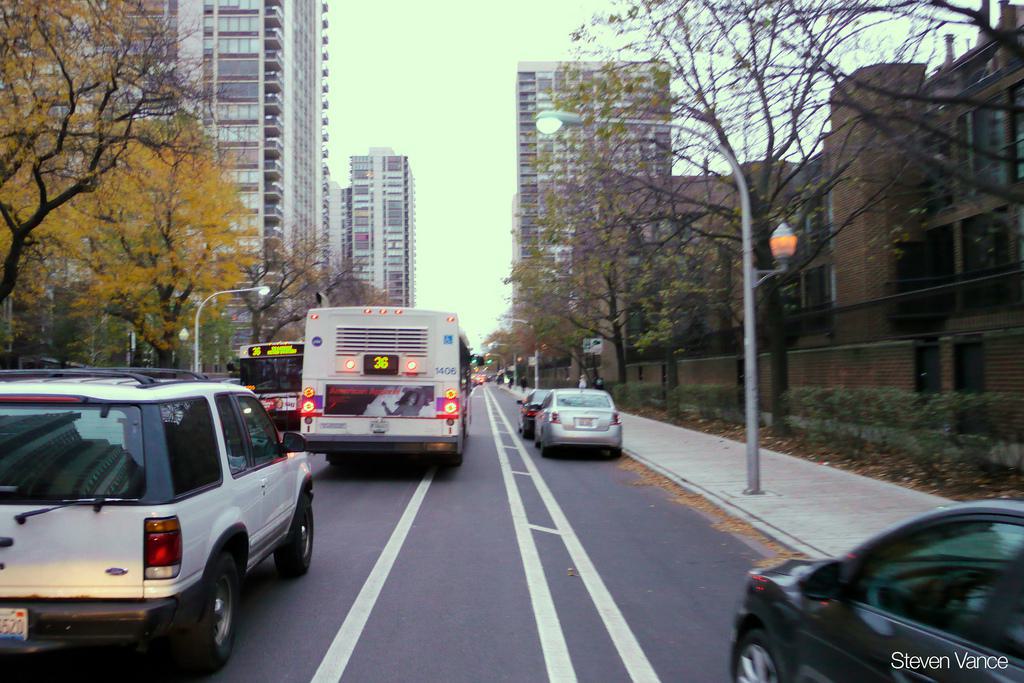}%
      }
    \end{adjustbox}
    &
    \begin{adjustbox}{valign=t}
      \begin{tcolorbox}[
        colback=white,
        colframe=green!70!black,
        boxrule=1pt,
        arc=2pt,
        left=2pt,
        right=2pt,
        top=2pt,
        bottom=2pt,
        width=\linewidth
      ]
      \footnotesize
      This image captures an urban street scene with tall residential buildings lining one side and leafy trees displaying autumn colors. The scene includes a city bus in the center of the frame, showing the number 36 on its indicator, and various other vehicles such as cars and SUVs. The road features a dedicated bike lane on the right, demarcated by white lines and identified by painted bicycle symbols. The overcast sky and the presence of a streetlight that is turned on suggest that this is either early morning or late afternoon. The photo appears to be taken from the perspective of a pedestrian or cyclist at street level, focused on capturing the flow of urban traffic.
      \end{tcolorbox}
    \end{adjustbox}
    &
    \begin{adjustbox}{valign=t}
      \begin{tcolorbox}[
        colback=white,
        colframe=red!70!black,
        boxrule=1pt,
        arc=2pt,
        left=2pt,
        right=2pt,
        top=2pt,
        bottom=2pt,
        width=\linewidth
      ]
      \footnotesize
      This image depicts an overcast day on an urban street lined with tall, modern office buildings. A \hl{blue public bus marked with the number 421} is at the forefront on the road, while a \hl{red bus} can be seen farther down the street. There is a white car on the left and \hl{traffic lights} are visible overhead, with a red light illuminated. The road has multiple lanes and a pedestrian zebra crossing in the foreground. There's also a \hl{traffic sign indicating no left turn} for motorcycles. Leafless trees suggest it may be winter or early spring. The overall scene appears to be calm with moderate traffic.
      \end{tcolorbox}
    \end{adjustbox}
    \\[1em]

    \midrule

    \begin{adjustbox}{valign=t}
      \makebox[\linewidth][c]{%
        \includegraphics[width=3cm,height=2.5cm]{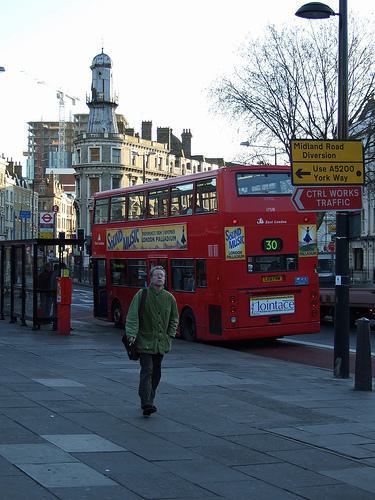}%
      }
    \end{adjustbox}
    &
    \begin{adjustbox}{valign=t}
      \begin{tcolorbox}[
        colback=white,
        colframe=green!70!black,
        boxrule=1pt,
        arc=2pt,
        left=2pt,
        right=2pt,
        top=2pt,
        bottom=2pt,
        width=\linewidth
      ]
      \footnotesize
      This image captures a British urban scene, highlighted by a classic red double-decker bus on the right, displaying route number 30. The bus has yellow text and advertisements printed on its sides. On the left, a pedestrian wearing a green jacket and carrying a bag seems to be briskly walking on the sidewalk. There's a yellow street sign indicating a diversion ahead. In the background, an ornate building towers with a clock at its apex under a clear blue sky. The street is flanked by various other buildings, likely a mix of residential and commercial structures, typical of a UK cityscape.
      \end{tcolorbox}
    \end{adjustbox}
    &
    \begin{adjustbox}{valign=t}
      \begin{tcolorbox}[
        colback=white,
        colframe=red!70!black,
        boxrule=1pt,
        arc=2pt,
        left=2pt,
        right=2pt,
        top=2pt,
        bottom=2pt,
        width=\linewidth
      ]
      \footnotesize
      This image captures a bustling urban scene, likely in London, with a red double-decker bus dominating the foreground, bearing the signage \hl{'Arriva' and a route number 176 to Penge via Elephant \& Castle and Forest Hill}. A person at a \hl{pedestrian crossing} is using a push-button signal post, while others wait by a bus stop shelter where someone points upwards. To the right, a \hl{classic red telephone box} is in use by an individual. In the background, neoclassical architecture suggests a historical district, with a dome-topped building visible in the distance. The street is lined with cars and traditional black iron fencing, contributing to a distinctly British urban landscape.
      \end{tcolorbox}
    \end{adjustbox}
    \\

    \bottomrule
  \end{tabular}
  }
  \caption{Qualitative comparison of image-text retrieval results between MulCLIP (middle column) and GOAL (right column). Borders are embedded to indicate correctness (green: correct; red: incorrect). \textbf{Yellow highlights} denote visual details missing from GOAL's retrieved image but correctly matched by MulCLIP.}
  \label{tab:image_2_text}
\end{table}

\end{document}